\documentclass{aastex}

\usepackage{graphicx}
\usepackage{emulateapj5}
\usepackage{times}



\def\chandra    {{\em Chandra}\/}

\def\xmm        {XMM-{\em Newton}\/}

\def\hst        {{\em HST}\/}
\def\spi        {{\em Spitzer}\/}
\def\gem        {{\em Gemini}\/}

\def\ga         {{ESO~137-001}\/}
\def\gab        {{ESO~137-002}\/}

\usepackage{mathptmx}

\begin{document}

\title{Spectacular X-ray tails, intracluster star formation and ULXs in A3627}

\author{
M.\ Sun,$^{\!}$\altaffilmark{1}
M.\ Donahue,$^{\!}$\altaffilmark{2}
E.\ Roediger,$^{\!}$\altaffilmark{3}
P.\ E.\ J.\ Nulsen,$^{\!}$\altaffilmark{4,5}
G.\ M.\ Voit,$^{\!}$\altaffilmark{2}
C.\ Sarazin,$^{\!}$\altaffilmark{1}
W.\ Forman,$^{\!}$\altaffilmark{4}
C.\ Jones$^{\!}$\altaffilmark{4}
}
\smallskip

\affil{\scriptsize 1) Department of Astronomy, University of Virginia, P.O. Box 400325, Charlottesville, VA 22901; msun@virginia.edu}
\affil{\scriptsize 2) Department of Physics and Astronomy, Michigan State University, East Lansing, MI 48824}
\affil{\scriptsize 3) Jacobs University Bremen, P. O. Box 750 561, 28725 Bremen, Germany}
\affil{\scriptsize 4) Harvard-Smithsonian Center for Astrophysics,
60 Garden St., Cambridge, MA 02138}
\affil{\scriptsize 5) University of Wollongong, NSW 2522, Australia, on leave} 	

\shorttitle{X-ray tails and Intracluster star-formation (III)}
\shortauthors{Sun et al.}

\begin{abstract}

We present the discovery of spectacular double X-ray tails associated with
\ga\ and a possibly heated X-ray tail associated with \gab, both late-type
galaxies in the closest rich cluster Abell~3627. A deep \chandra\
observation of \ga\ allows us for the first time to examine the
spatial and spectral properties of such X-ray tails in detail. Besides the
known bright tail that extends to $\sim$ 80 kpc from \ga, a fainter and
narrower secondary tail with a similar length was surprisingly revealed,
as well as some intriguing substructures in the main tail. There is little
temperature variation along both tails. The widths of the secondary tail
and the greater part of the main tail also remain nearly constant with
the distance from the galaxy. All these results challenge the current
simulations. The \chandra\ data also reveal 19 X-ray point sources
around the X-ray tails. We identified six X-ray point sources as candidates
of intracluster ULXs with $L_{\rm 0.3 - 10 keV}$ of up to 2.5$\times10^{40}$
erg s$^{-1}$. \gem\ spectra of intracluster HII regions downstream of \ga\
are also presented, as well as the velocity map of these HII regions that
shows the imprint of \ga's disk rotation. For the first time, we unambiguously
know that active star formation can happen in the cold ISM stripped by ICM
ram pressure and it may contribute a significant amount of the intracluster
light. We also report the discovery of a 40 kpc X-ray tail of another
late-type galaxy in A3627, \gab. Its X-ray tail seems hot,
$\sim$ 2 keV (compared to $\sim$ 0.8 keV for \ga's tails). The H$\alpha$
data for \gab\ are also presented. We conclude that the high pressure
environment around these two galaxies is important for their bright X-ray
tails and the intracluster star formation. The soft X-ray tails can reveal
a great deal of the thermal history of the stripped cold ISM in mixing with
the hot ICM, which is discussed along with intracluster star formation.

\end{abstract}


\section{Introduction}

The intracluster medium (ICM) has long been proposed to play a vital role in
galaxy evolution in clusters, through ram pressure and turbulent/viscous
stripping of the galactic cold gas (e.g., Gunn \& Gott 1972; Nulsen 1982;
Quilis et al. 2000). As the halo gas and the cold interstellar medium (ISM) is depleted in the
stripping process, the galactic star formation will eventually be shut down
and blue disk galaxies may turn into red galaxies (e.g., Quilis et al. 2000).
The removal of the cold ISM also affects the accretion history of the central SMBH.
Stripping of the ISM of the disk galaxies in clusters has been extensively
studied in simulations recently, with better resolution and more physics included
(e.g., Abadi et al. 1999; Stevens et al. 1999; Quilis et al. 2000;
Schulz \& Struck 2001; Bekki \& Couch 2003; Roediger \& Hensler 2005;
Kapferer et al. 2009). These simulations show that stripping has a significant
impact on galaxy evolution (e.g., disk truncation, formation of flocculent arms,
build-up of a central bulge and enhanced star formation in the inner disk at the
early stage of the interaction).

Besides the impact on galaxy evolution, another significant question related
to stripping is the evolution of the stripped ISM. After the cold ISM is removed
from the galaxy, the general wisdom is that the stripped gas is mixed with the
ICM through evaporation eventually. However, it is now known that a fraction of
the stripped ISM turns into new stars in the
galactic halo or the intracluster space, as revealed from observations
(e.g., Sun et al. 2007b, S07 hereafter) and simulations (e.g., Vollmer et al. 2001b;
Schulz \& Struck 2001; Kronberger et al. 2008; Kapferer et al. 2009).
The stripped cold ISM, if it can survive long enough to
reach the cluster center, can effectively heat the cluster core via
ram-pressure drag, which makes gravitational heating by accretion a possible way
to offset cooling (Dekel \& Birnboim 2008 argued for 10$^{5}$ - 10$^{8}$ M$_{\odot}$ clumps).
Current HI observations of cluster spiral galaxies still fail to detect most
of the HI gas missing from HI-deficient spiral galaxies in the intracluster space
(e.g., Vollmer \& Huchtmeier 2007), which implies that the bulk of the stripped
HI gas has been heated out of the cold phase.
However, little is known about the details of mixing, as the actual strength of heat
conductivity and viscosity is poorly known. How quickly is the stripped cold
ISM heated and mixed with the ICM? What observational signature will evaporation and
mixing produce? What fraction of the stripped cold ISM turns into new stars?
The mixing of the stripped cold ISM with the hot ICM will produce multi-phase
gas. Depending on the poorly known details of mixing, prominent soft X-ray
emission may be produced, as well as H$\alpha$ emission. In this ``unified model''
for tails of cluster late-type galaxies, X-ray and H$\alpha$ tails are simply
manifestations of the cold ISM tail in the mixing process. Therefore, one can
better understand mixing and the relevant micro-physics from multi-wavelength
data, e.g., the amount of missing HI gas and the amount of the stripped gas in
hotter phases. Obviously, a central problem is the energy transfer in the multi-phase
gas, which is also a significant question for large cool cores in clusters. Moreover,
the wake behind the galaxy produced by stripping provides a way to constrain the ICM
viscosity (e.g., Sun et al. 2006, S06 hereafter;
Roediger \& Br$\ddot{\rm u}$ggen 2008a).

Various dynamic features, e.g., bow shocks, tails and vortices, can be produced
in stripping (e.g., Stevens et al. 1999; Schulz \& Struck 2001; Roediger et al. 2006;
Roediger \& Br$\ddot{\rm u}$ggen 2008b, RB08 hereafter; Tonnesen \& Bryan 2009).
Observational evidence of stripping of cluster late-type galaxies is
present in HI and H$\alpha$ observations, either through HI deficiency or tails
(e.g., Giovanelli \& Haynes 1985; Gavazzi et al. 2001b; Kenney et al. 2004;
Oosterloo \& van Gorkom 2005; Chung et al. 2007; Yagi et al. 2007; S07).
X-ray tails of late-type cluster galaxies have only been detected recently:
C153 in A2125 at $z$=0.253 (Wang et al. 2004), and UGC~6697 in A1367 at $z$=0.022
(Sun \& Vikhlinin 2005). However, there are only $\sim$ 60 counts from C153 so
the X-ray extension is not unambiguous (Wang et al. 2004). The \xmm\ data show
that the X-ray tail of UGC~6697 may extend to $\sim$ 90 kpc from the nucleus
but most of the X-ray emission is within the galaxy. Moreover,
the galaxy has a peculiar morphology and a tidal tail. Its complex
kinematic behavior, revealed from the velocity map, implies the existence
of a second galaxy hidden behind the main galaxy (Gavazzi et al. 2001a).
Thus, UGC~6697 is a very complicated system where tidal interaction is
important. There is also a wide tail (67 kpc$\times$90 kpc) of the spiral NGC~6872
in the 0.5 keV Pavo group (Machacek et al. 2005), but it is uncertain how it was
formed as the tail terminates on the dominant early-type galaxy of the group.
The brightest known X-ray tail behind a cluster late-type galaxy is \ga\ in the
closest rich cluster A3627 (S06), with $\sim$ 80\% of the X-ray emission beyond the
galactic halo. One should be aware that strong X-ray tails of late-type
galaxies are rare (e.g., none in Virgo, see Section 7.1).
We have done a blind search of strong X-ray tails in nearby clusters ($z<0.06$) by extending
our previous \chandra\ work (Sun et al. 2007a) to the current \xmm\ and \chandra\ archives.
Despite extensive cluster data in the archives, e.g., mosaic fields of $\sim$ 2 deg$^{2}$
around the Perseus and Coma clusters with \xmm, only two more X-ray tails of cluster
late-type galaxies were found. One is also in A3627 (\gab) and will be discussed in
this paper. The other one is NGC~4848 in the Coma cluster (Finoguenov et al. 2004)
that was recently observed for 29 ks by \chandra. However, both X-ray tails are
shorter (40 - 50 kpc) and 2.4 - 5.5 times fainter than \ga's tail. Thus, the proximity
of \ga, the high flux and the large length of its X-ray tail makes it the best
target for detailed analysis and comparison with simulations. One should not
confuse X-ray tails of early-type galaxies with those of late-type galaxies.
Early-type galaxies in clusters have abundant X-ray ISM but little or no cold ISM.
There are some X-ray tails of early-type galaxies reported (e.g.,
Machacek et al. 2006; Sun et al. 2007; Randall et al. 2009). Their X-ray tails
are composed of the stripped hot ISM and do not co-exist with cold gas.

A3627 is the closest massive cluster ($z$=0.0163, $\sigma_{\rm radial}$
= 925 km/s and $kT$=6 keV) rivaling Coma and Perseus in mass and galaxy content
(Kraan-Korteweg et al. 1996; Woudt et al. 2008). It is also the sixth brightest
cluster in RASS and the second brightest non-cool-core cluster after Coma
(B$\ddot{\rm o}$hringer et al. 1996). \ga\ is a blue emission-line galaxy
(Woudt et al. 2004) that is only $\sim$ 180 kpc from the cluster's X-ray peak
in projection. Its radial velocity (4680$\pm$71 km~s$^{-1}$, Woudt et al. 2004)
is close to the average velocity in A3627 (4871$\pm$54 km~s$^{-1}$, Woudt et al. 2008)
so most of its motion is probably in the plane of sky. If \ga\ is on a radial
orbit, its real distance to the cluster center should be close to the
projected distance. There is no sign of a galaxy merger and no strong tidal
features around \ga. S06 found a long X-ray tail behind \ga, in both
\chandra\ (14 ks) and \xmm\ data (MOS: 18 ks, PN: 12 ks). The tail extends
to at least 70 kpc from the galaxy with a length-to-width ratio of $\sim$ 10.
The X-ray tail is luminous ($\sim$ 10$^{41}$ erg s$^{-1}$), with an X-ray gas mass
of $\sim 10^{9}$ M$_{\odot}$. S06 interpreted the tail as the stripped ISM of
\ga\ mixed with the hot ICM. The \chandra\ data also reveal three hard X-ray point
sources ($L_{X} \sim 10^{40}$ erg s$^{-1}$) along the tail, and the possibility
of all of them being background active galactic nuclei (AGNs) is very small. S06 suggested
that some of them may be ultra-luminous X-ray sources (ULXs) born from active
star formation in the tail. S07 further discovered a 40 kpc H$\alpha$ tail and
over 30 emission-line regions downstream of \ga, up to 40 kpc from the galaxy.
S07 concluded that they are giant HII regions in the halo downstream
or in intracluster space.
Sivanandam et al. (2010) observed \ga\ with IRAC and IRS on \spi.
A warm ($\sim$ 160 K) molecular hydrogen tail was detected to at least 20 kpc
from the galaxy from the IRS data, at the same position as the bright X-ray
and H$\alpha$ tail. The total mass of the warm H$_{2}$ gas is
$\sim 2.5\times10^{7}$ M$_{\odot}$. As the IRS fields only cover the front part of the
X-ray tail, the above warm H$_{2}$ gas mass is only a lower limit.
8 $\mu$m PAH emission is also detected from the bright HII regions identified
by S07. The \spi\ results may imply a large reservoir of colder gas than the observed
warm gas downstream of \ga, coexisting with the hot ICM ($\sim 7\times10^{7}$ K) and the
H$\alpha$ emitting gas ($\sim 10^{4}$ K).

Obvious follow-ups include deep \chandra\ exposures and optical spectroscopic
observations of the candidate HII regions. In this paper, we present the results from
our deep \chandra\ observation of \ga\ and \gem\ GMOS spectroscopic
observations. The plan of this paper is as follows: The \chandra\ and \gem\
observations and the data analyses are presented in Section 2. In Section 3, we
discuss the X-ray tails of \ga. Section 4 is on the properties of the surrounding ICM. 
Section 5 discusses the \chandra\ point sources and the intracluster HII regions.
We also found another 40 kpc X-ray tail in A3627,
associated with \gab, which is discussed in Section 6. The results are discussed
in Section 7 and Section 8 contains the summary.
We adopt a cluster redshift of 0.0163 for A3627 (Woudt et al. 2008).
Assuming H$_{0}$ = 71 km s$^{-1}$ Mpc$^{-1}$, $\Omega$$_{\rm M}$=0.27,
and $\Omega_{\rm \Lambda}$=0.73, the luminosity distance is 69.6 Mpc,
and 1$''$=0.327 kpc. 

\section{New Observations of \ga}

\subsection{The Deep Chandra Data}

The observation of \ga\ was performed with the Advanced CCD Imaging
Spectrometer (ACIS) on June 13 - 15, 2008 (obsID: 9518). Standard \chandra\ data
analysis was performed which includes the corrections for the slow gain change
and charge transfer inefficiency (for both the FI and BI chips). We examined
the light curve from the part of the S3 chip with weaker ICM emission.
No flares of particle background were found. There are also no flares in
the light curve of the I2+I3 chips. The effective exposures are 138.2 ks for the BI
chips and 140.0 ks for the FI chips. This new \chandra\ exposure is 18 times deeper
than the old one. About 4100 net counts from the thermal gas were collected in the
0.5 - 3 keV band. We corrected for the ACIS low-energy quantum efficiency (QE)
degradation due to the contamination on ACIS's optical blocking filter,
which increases with time and varies with position. The dead area
effect on the FI chips, caused by cosmic rays, has also been corrected.
We used CIAO3.4 for the data analysis. The calibration files used
correspond to \chandra\ calibration database (CALDB) 3.5.3 from the
\chandra\ X-ray Center. We used the new HRMA on-axis Effective Area file,
``hrmaD1996-12-20axeffaN0008.fits'', which improves the accuracy of
the temperature measurement at $kT >$ 4 keV. The remaining uncertainty in
the source flux (up to $\sim$ 10\%) would not much affect our conclusions in this work.
In the spectral analysis, a lower energy cutoff of 0.4 keV (for the BI data)
and 0.5 keV (for the FI data) is used to minimize the effects of calibration
uncertainties at low energy. The solar photospheric abundance table by
Anders \& Grevesse (1989) is used in the spectral fits. Uncertainties
quoted in this paper are 1 $\sigma$. We used Cash statistics in this work.
We adopted an absorption column density of 1.73$\times10^{21}$ cm$^{-2}$ from the
Leiden/Argentine/Bonn HI survey (Kalberla et al. 2005). If we choose to fit
the absorption column from the \chandra\ spectra (with the PHABS model in XSPEC),
the best fits are always
consistent with the above value. This absorption column is lower than
the previous value from Dickey \& Lockman (1990), 2.0$\times10^{21}$ cm$^{-2}$,
which had been used in previous work (e.g., B$\ddot{\rm o}$hringer et al. 1996; S06).

\subsection{The Gemini Data}

We obtained optical spectra of the emission-line objects identified by S07 
with GMOS on the \gem\ south (program ID: 2008A-Q50). 
Because the field is crowded with interesting sources, three masks had to be
used with some sources covered in more than one mask. Observations with the
first mask were done on May 3, 2008 (airmass: 1.18 - 1.22, seeing: 
$\sim 0.9''$). Two exposures (20 minutes each) were taken, with a central
wavelength difference of 60 \AA\ in order to cover the inter-chip gaps.
Observations with the second mask and the third mask were
taken on June 26 and June 27, 2008 respectively (airmass: 1.17, seeing: 
$\sim 2''$ for June 26; airmass: 1.18, seeing: $\sim 1.3''$ for June 27).
Two exposures (10 minutes each) were taken on each night.
All three nights were photometric. The R400 grating + GG455 filter were used,
which gives a free wavelength range of $\sim$ 4500 - 8000 \AA.
We used 1$''$ slits for all sources.
As queued observations, we used the calibration data of LTT4364 taken on the 
photometric night of May 27, 2008 for the relative flux calibration.

The GMOS data were reduced with the Gemini IRAF Package (version 1.9.1).
We generally followed the standard procedures. Besides the accompanying CuAr
arc data, we also used the night sky lines for the wavelength calibration.
The latter procedure is the key to derive consistent velocities for the same
source from different exposures. In this paper, we present only the velocity
information and briefly discuss the nature of these sources (HII regions).
The detailed properties of these HII regions will be discussed in a future
paper.

\section{X-ray Tails of \ga}

The 0.6 - 2.0 keV \chandra\ count image is shown in Figure 1, with point
sources removed. The image was smoothed to enhance extended low surface
brightness features. Compared with images from short \chandra\ and \xmm\
exposures, a secondary tail is clearly visible to the south of the main
tail. Some substructures are also revealed, including a ``protrusion'' from the
main tail and two sharp bends in the main tail. Also in Figure 1, the
X-ray contours from the adaptively smoothed image (exposure corrected, not
the one in the left panel) are overlaid on the H$\alpha$+continuum image from
the Southern Observatory for Astrophysical Research (SOAR).
We also show the \xmm\ image of the cluster and a composite color image
of \ga's tails (Figure 2). In the following sections, we discuss the spatial
and spectral properties of \ga's tails.

\subsection{Spatial Structure}

The observed soft X-ray emission of the tails most likely comes from the
interfaces between the hot ICM and the cold stripped ISM. Although there
are no data for the cold atomic and molecular gas (note the \spi\
data on the PAH emission and the warm molecular H$_{2}$, Sivanandam et al. 2010), 
the X-ray surface brightness of the tails reveals a great deal about
the stripped ISM. The surface brightness of the tails is quantitatively
examined. We first derived the 0.6 - 2 keV surface brightness profiles
across some parts of the tails and the leading edge. As shown in Figure 3,
both tails are narrow but highly significant features above the local
background. Even the ``protrusion'' is a 4.1 $\sigma$ feature. 
The leading edge (or the contact discontinuity) is very sharp, corresponding
to the H$\alpha$ edge (Figure 4). The gap between two tails has a similar
surface brightness as the local background (Fig. 3b and 3c).
For the front part of the gap, the surface brightness
is (6.26$\pm$0.30)$\times10^{-5}$ cnts s$^{-1}$ kpc$^{-2}$, while the
local background is (5.88$\pm$0.27)$\times10^{-5}$ cnts s$^{-1}$ kpc$^{-2}$.
The low surface brightness of the gap has implications for the 3D structure
of the stripped ISM tail. One possibility for the observed double tails is
the projection effect of one broad cold ISM tail. According to this
interpretation, the 3D structure of the soft X-ray emission might be a
cylindrical shell that produces double tails in the projected surface
brightness. As shown in Figure 3, such a configuration fails to simultaneously
account for both the low surface brightness of
the gap and the observed widths of two tails. The observed emission level
between tails is similar to that outside tails. Basically, the width
of the cylindrical shell required to make the central emission dim
also makes each peak unacceptably narrow in relation to the observed profiles.
Thus, the two X-ray tails are most likely detached. This is
also supported by the \spi\ data (Sivanandam et al. 2010). The 8 $\mu$m PAH emission
only appears in the first 20 kpc of the main tail. The IRS data only
cover a small field downstream but no enhanced warm H$_{2}$ emission
is present at the front part of the gap. Eventually we need HI and CO data
to better reveal the distribution of the cold ISM downstream of \ga.

In principle, the surface brightness profiles along the tails reflect the
stripping history. As shown in Figure 5, both tails have wiggles and substructures
along the (time) axis. However, the unknown conversion from the X-ray
emission to the mass of the stripped ISM (filling factor, detail of the
emission spectra, etc.) makes this difficult with only the X-ray data.
A more complete picture can be obtained with the HI and CO data. From
Figure 5, the total linear length of the whole tail system is at least
4.2$'$ or 80 kpc.
We also measured the widths of the main tail in four segments (before
sharp bends) and the widths of the secondary tail in two segments.
The model is composed of a Gaussian and a flat local background. As shown in Figure 6,
the widths of both tails do not increase with the distance from the galaxy.
This conflicts with simulations
(e.g., RB08; Kapferer et al. 2009), as further discussed in Section 7.2.
If \ga\ is moving supersonically, a bow shock may form. The local
sound speed is 1263 ($kT$ / 6 keV)$^{1/2}$ km/s. For a Mach number of 1.1 - 3,
the predicted position of the bow shock is 20$'' - 2''$ from the leading edge
(or the contact discontinuity). No such shock front is detected in
the \chandra\ exposure.
The enhanced surface brightness would be much weaker in projection
on the sky, especially if \ga\ is farther from the cluster center
than the projected position.

\subsection{Spectral Properties}

With this deep \chandra\ exposure, for the first time spatially-resolved
spectroscopy can be done for X-ray stripping tails of a cluster late-type galaxy.
We measured gas temperatures in five regions of the main tail and two regions
of the secondary tail (Figure 7). The background is from the local source-free regions.
A single APEC model with Galactic absorption
was used. We emphasize that the measured temperatures are only spectroscopic
temperatures as the intrinsically multi-phase gas is fitted with a single-$kT$ model.
In fact, a hard X-ray excess generally exists in the spectral fits, which implies
the existence of hotter components. However with limited statistics, we elected to
keep the spectral model simple.
As shown in Figure 7, the variations of spectroscopic temperatures along both tails are
small. In fact, both tails have consistent temperatures.

Since temperatures derived from individual regions are similar, the total spectrum
of two tails (still excluding the ``head'' region) was also examined with several
models (Figure 8 and Table 1). This observation collects 3444 net counts from the
thermal gas of the two tails in the 0.5 - 3 keV band. While the fit with a single-$kT$ model
is acceptable, the addition of a second component (either another thermal component
or a power-law component) produces better fits. We also tried the XSPEC model
CEMEKL, which allows a power-law distribution of the emission measure for multi-$kT$ gas.
We consider the spectral results shown in Table 1 as evidence for multi-phase gas,
especially as the single-$kT$ model produces a very low abundance.
Adding another or more thermal components can increase the abundance to the solar
value. Nevertheless, the double APEC or the CEMEKL models are no more than
a phenomenological way to describe the data. It is still unclear what emergent
spectrum mixing will produce. There are more complications, like the
non-equilibrium ionization effect and the intrinsic absorption. Generally,
ionization equilibrium requires the collisional ionization parameter
$n_{\rm e} t > 10^{12}$ cm$^{-3}$ s. For $n_{\rm e} = 10^{-2} - 10^{-3}$
cm$^{-3}$, $t >$ 3 - 32 Myr, which is comparable to the age of the X-ray tails.

The X-ray gas mass of the tails can be estimated. The main uncertainties are
the geometry (or the filling factor) of the X-ray emitting gas and the
abundance of the gas. The filling factor of the soft X-ray emitting gas
depends on the mass and configuration of the stripped cold ISM gas in the tails.
The soft X-rays may originate from the interfaces between the ambient hot ICM and
the stripped cold ISM. A fragmented cold ISM tail has larger surface area
so the filling factor of the soft X-ray emitting gas would be larger.
Both uncertainties can be constrained by the assumed pressure equilibrium
with the ambient ICM. In the following, we present the estimates from the
one-$kT$ fit, the two-$kT$ fit and the CEMEKL fit. We assume both the main tail and the
secondary tail are cylinders with a length of 80 kpc and radii of 3.9 kpc
and 3.4 kpc respectively. The ``head'' and ``protrusion'' regions are not included. 
With the result from the one-$kT$ fit (the first row of Table 1), the total X-ray
gas mass is 2.0$\times10^{9}$ $f^{1/2}$ M$_{\odot}$, where $f$ is the filling
factor of the soft X-ray emitting gas. If the estimate is done on the same individual
regions as the spectral analysis (Figure 7), the total gas mass only
decreases by 3\%. The electron density ranges from 0.0080 $f^{-1/2}$ cm$^{-3}$
to 0.017 $f^{-1/2}$ cm$^{-3}$ along the tails. However, the abundance of the X-ray gas is
too low for the one-$kT$ fit, which results in gas densities that are too high.
The ambient ICM electron density and temperature in the projected position of \ga's tails is
$\sim 1.3\times10^{-3}$ cm$^{-3}$ and 6.3 keV respectively. This puts an upper limit of
0.010 cm$^{-3}$ on the electron density of the tail gas (for a temperature of 0.8 keV)
if pressure equilibrium is assumed. Thus, the one-$kT$ model is not sufficient.

Do the two-$kT$ fit and the CEMEKL fit produce more reasonable results? We first used
the result in the fourth row of Table 1, where the abundance of the lower temperature
component is fixed at the solar value and the abundance of the higher temperature
component is fixed at the value of the surrounding ICM. Assuming pressure
equilibrium between these two components, it is found that the hotter
component occupies 97\% of the volume of the soft X-ray emitting gas.
The electron density of the cool and hot components is 0.018 $f^{-1/2}$ cm$^{-3}$
and 0.0071 $f^{-1/2}$ cm$^{-3}$ respectively. The total X-ray gas mass is
1.4$\times10^{9}$ $f^{1/2}$ M$_{\odot}$. However, the gas pressure in the
tail is $\sim$ 50\% higher than the maximum ICM pressure of the surroundings.
This problem cannot be solved by increasing the abundance of the hotter
component in the spectral fit, which will further increase the ISM
pressure. We then tried the CEMEKL fit (the eighth row of Table 1). However,
the X-ray tail is at least 3.2 times over-pressured to the surrounding ICM,
under an isobaric condition. The total X-ray gas mass is 1.1$\times10^{9}$ M$_{\odot}$.
Therefore, either these models are not adequate, or there is dynamic pressure
(e.g., turbulence) besides the thermal pressure.
Without better understanding of the tail spectrum and the contribution
of the cold ISM in the tails, it is difficult to much improve the above
estimates. We summarize \ga's properties from the current data in Table 2,
along with those of \gab\ (Section 6).

\section{Properties of the Surrounding ICM}

We also examined the spectral properties of the ICM around \ga's tails.
We followed the same approach as used in Sun (2009) for ESO~137-006 to
determine the local soft X-ray background that is dominated by the Galactic
emission (note the Galactic latitude of -7 deg).
With the soft X-ray background determined, ICM temperatures were determined
in 12 regions around \ga's X-ray tails (Figure 7). The projected ICM temperature
is about 6 keV around \ga's tails. Region \#8 has the lowest
temperature, which implies the presence of a faint ISM component in that
region. Indeed, the 0.6 - 2 keV surface brightness at region \#8 is a bit higher
than that in regions \#11 and \#13, (5.53$\pm$0.13)$\times10^{-5}$ cnts s$^{-1}$
kpc$^{-2}$ vs. (5.23$\pm$0.10)$\times10^{-5}$ cnts s$^{-1}$ kpc$^{-2}$.
Region \#7 may include some small ``protrusions'' from both tails (Figure 1),
which can explain its relatively lower temperature than regions \#2 - \#6.
The ICM abundance can also be determined from a simultaneous fit to the spectra
of all regions, 0.20$\pm$0.06 solar.
The electron density of the surrounding ICM is 0.0010 cm$^{-3}$ - 0.0014 cm$^{-3}$
in the projected positions of the tails (B$\ddot{\rm o}$hringer et al. 1996).
The total ICM thermal pressure is 1.8$\times10^{-11}$ dyn cm$^{-2}$ for $n_{\rm e}$ =
10$^{-3}$ cm$^{-3}$ and $kT$=6 keV. The ram pressure is 4.4$\times10^{-11}$
($v_{\rm gal}$/1500 km/s)$^{2}$ dyn cm$^{-2}$ for $n_{\rm e}$ = 10$^{-3}$ cm$^{-3}$.
Both are high enough to strip most part of disk gas, except for the very
core (e.g., Roediger \& Br$\ddot{\rm u}$ggen 2005).

\section{Intracluster HII Regions and Ultraluminous X-ray Sources}

\subsection{Intracluster HII Regions and the \gem\ Spectra}

S07 identified 29 strong candidate HII regions. Six more sources were
also identified as possible HII regions. The \gem\ GMOS observations were
able to cover thirty-three of them (solid blue squares in Figure 9).
Only two sources (dashed blue squares in Figure 9) were not observed.
Both of them are strong candidates but are faint and close to
\ga. The GMOS spectra unambiguously confirm the association of these
33 sources (27 strong candidates + 6 weaker candidates in S07)
with \ga\ as their velocities are all within 200 km/s of
\ga's. The heliocentric velocity map of these 33 HII regions is shown in Figure 10.
In Figure 10, we also plot their velocities vs. their distances to the major
axis of the galaxy or the middle line of the bow-like front of HII
regions. We used a system velocity of 4660 km/s for \ga, which is the
weighted average from the velocities derived by Woudt et al. (2004) and S07.
The velocities of these HII regions span a small range ($\leq$ 300 km/s)
with a dispersion of 54$^{+12}_{-10}$ km/s. In fact, this small velocity
range is comparable to the radial rotation velocity of the residual H$\alpha$
nebula at the center of \ga, $\sim$ 80 km/s from our SOAR GOODMAN spectra
that will be presented in a future paper. After ISM is removed from the
galactic disk, it is subject to turbulence in the wake. Combined with
the residual rotational motion, the range of the radial velocities of
stripped ISM clouds should become substantially larger than the range of the
rotation velocity (e.g., RB08). However, that is not what we observed
with the velocities of HII regions as tracers. In fact as shown in Figure 10,
the imprint of the rotation velocity pattern is still present in the velocity
map of these HII regions.

Some example spectra of these HII regions are shown in Figure 11, including
several sources that are most distant from \ga. The low
[OI] $\lambda$6300 / H$\alpha$ and [NII] $\lambda$6584 / H$\alpha$ ratios
are typical of giant HII regions. The detailed properties of these HII
regions (e.g., line equivalent width, line ratios,
star-formation rate, age, metallicity and total mass)
will be presented in a future paper with the results from our \hst\ data.

\subsection{\chandra\ Point Sources}

The deep \chandra\ exposure also reveals many point sources, as shown in
Figure 9. We focus on X-ray point sources in \ga\ or around its X-ray tails
(within the two elliptical regions in Figure 9). Nineteen point sources are detected.
Their positions and fluxes are listed in Table 3.
The 0.5 - 2 keV limiting flux is $\sim 3.5\times10^{-16}$ erg s$^{-1}$
in the region of interest (with eight counts in the 0.5-2 keV band).
The expected X-ray source density above this limiting flux is
$\sim 1500$ per deg$^{2}$ (Kim et al. 2007). For the two elliptical regions
(9.66 arcmin$^{2}$), the expected number of background sources is four,
while 19 point sources are detected. The contribution of Galactic sources
should not be neglected as \ga\ is located at a low Galactic latitude (-7 deg).
Indeed, six \chandra\ point sources are positionally coincident with bright
star-like objects. However, an excess of X-ray point sources still exists,
especially in the immediate downstream region of \ga\ where most HII regions are.
We can divide these point sources into three groups, sources 1-3 that are upstream
of the galaxy, sources 4-12 that are around the disk plane or downstream of
the galaxy (but still within the 15 kpc radius halo in projection) and sources
13-19 that are unbound and near the X-ray tails. The latter two groups are
of most interest here. The region immediately downstream of the galaxy
(but still within the galactic halo) has nine X-ray sources, while only 0.5
are predicted from the Log N - Log S relation ($\sim$ 1.1 arcmin$^{2}$).
The unbound tail region has seven X-ray sources, while 3.3 sources
are expected from the Log N - Log S relation ($\sim$ 7.9 arcmin$^{2}$).
Three sources are likely of stellar origin (14, 16 and 17), while source 15
is close to a confirmed HII region. If we take 3 stars in the unbound
tail region as the number density of stars, only 0.4 stars are expected in the
immediate downstream region. In fact, only source 7 coincides with a bright star.
Thus, the excess of X-ray point sources in the immediate downstream region
is significant.

We examined the optical counterparts of these X-ray point sources and list
the results in Table 3. The SOAR images ($B$, $I$ and H$\alpha$ bands) that are
presented in S07 are used. With seven sources (across the S3 chip) detected
in both the X-rays and the optical images, the astrometry can be
determined to $\sim 0.3''$. There are four
\chandra\ sources (8, 10, 11 and 15) that are within 2$''$ (or 0.65 kpc) of the
confirmed HII regions and have no other optical counterparts in the SOAR images.
We consider them strong candidates for intracluster X-ray binaries, while
a 0.5 kpc offset can be produced by a 100 km/s kick velocity in 5 Myr.
Sources 9 and 12 are 0.85 - 1.3 kpc from confirmed HII regions and do not
fall on any optical sources. They are also candidates for intracluster X-ray binaries.
If associated with \ga, their 0.3 - 10 keV luminosities are
$0.92 - 25 \times 10^{39}$ erg s$^{-1}$ (Table 3), which makes them
intracluster ULXs.
The X-ray spectra of the bright point sources were also examined (Table 3).
It is true that the chance match between HII regions and X-ray point sources
is high in the immediate downstream region. There is a $\sim$ 52\% chance that
an X-ray source will be within 2$''$ of an HII region there. However, the
excess of X-ray sources is also significant there.

As ULXs are often found around young massive star clusters, it may not
be a surprise to have intracluster ULXs around intracluster massive star clusters.
We can further examine the relation between the X-ray luminosities of these
putative ULXs and the total star-formation rate of the HII regions.
The total 2-10 keV luminosity of sources 8, 10, 11 and 15 is 6.5$\times10^{39}$
erg s$^{-1}$. If sources 9 and 12 are added, the total luminosity is
2.7$\times10^{40}$ erg s$^{-1}$. From S07, the total SFR of these 33 HII
regions is $\sim$ 0.7 M$_{\odot}$/yr, assuming an intrinsic extinction
of 1 mag for the H$\alpha$ emission. This SFR and the lower value of the
X-ray luminosity fall well on the relation derived by Grimm et al. (2003).
Even for the higher value of the X-ray luminosity (when sources 9 and 12
are included), agreement can be achieved if the intrinsic extinction
is higher in these HII regions and the SFR in the past was higher.

\section{\gab}

Another X-ray tail of a late-type galaxy was serendipitously found in our
ACIS-S observation of ESO~137-006 (the BCG of A3627, Sun 2009).
It is associated with \gab, a redder and more massive disk galaxy than \ga\ (Table 2).
\gab\ is only $\sim$ 110 kpc from the cluster's X-ray peak in projection but
its radial velocity is $\sim$ 900 km/s larger than A3627's.
The tail is near the corner of the I3 chip that is 10.8$'$ from the optical axis
of the observation (Figure 12 and 13). \gab\ was also covered by one of
our previous \chandra\ ACIS-I observations of A3627 (obsID: 4958, PI: Jones)
but the exposure is too short (14 ks) to reveal the X-ray tail.
We summarize the properties of \gab's tail in this paper, mainly for the purpose of
comparison with \ga's tails.

\subsection{The \chandra\ and SOAR Data}

The details of the \chandra\ observation of ESO~137-006 (obsID 8178) and of the
data analysis were presented in Section 5 of Sun (2009). The effective exposure
was 57.3 ks for the I2 chip. A significant X-ray tail behind \gab\ is revealed,
which in fact can also be seen in the \xmm\ data (Figure 2).

Being aware of its X-ray tail, we observed \gab\ ($z$=0.0191) with the 4.1-m
SOAR telescope on Cerro Pachon
on Apr. 18, 2008 (UT). The observations were made with the SOAR
Optical Imager (SOI), which covers a 5.2$' \times 5.2'$ field of view.
The night was photometric and seeing was very good, 0.45$''$-0.65$''$.
The airmass was 1.16 - 1.20.
We took three 800-sec exposures with the CTIO 6693/76 filter (H$\alpha$ band)
and two 600-sec exposures with the CTIO 6606/75 filter (H$\alpha_{\rm off}$ band)
\footnote{http://www.ctio.noao.edu/instruments/filters/index.html}.
Both narrow-band filters are 2 inches $\times$ 2 inches, so the
non-vignetted field is only $\sim 3.6' \times 3.5'$. Dithers of
15$'' - 30''$ were made between exposures.
Each image was reduced using the standard procedures with the IRAF
MSCRED package. The pixels were binned 2$\times2$, for a scale of
0.154$''$ per pixel. Dome flats were used. The US Naval
Observatory (USNO) A2.0 catalog was used for the WCS alignment, which
was done with WCSTools. The spectrophotometric standards were LTT~4364 and
EG~274 (Hamuy et al. 1994).

\subsection{\gab's Properties and Its X-ray Tail}

\gab\ is a disk galaxy with a dust lane across its bulge (Figure 13). It is
a more massive galaxy than \ga\ and has a larger radial velocity component
than \ga\ (Table 2). \gab\ is 120 kpc (or 6.15$'$) from \ga\ in the plane of sky.
Assuming the same ratios of the halo mass
to the stellar mass, the tidal radius [$r_{\rm t} \sim (m/3M)^{1/3} D$] is
0.32 -0.41 $D$, where $D$ is the distance between the two galaxies.
This value is much larger than the tidal truncation radius of \ga\ ($\sim$ 15 kpc,
truncated by the cluster potential) estimated by S07. Even if the two galaxies are
only separated by their projected distance, \ga's gravity is over 6 times \gab's
within 15 kpc radius of \ga's nucleus. Given the large difference of their radial
velocities ($\sim$ 1080 km/s), their actual spatial separation should be larger
than the projected distance, especially at the time when stripping began.
Thus, the tidal interaction between two galaxies should be small. There are indeed
no tidal features upstream of \ga\ and around \gab. Their X-ray tails are
also pointing away from each other.

\gab's X-ray tail can be traced to at least 40 kpc from the nucleus. Since it is
10.6$'$ off-axis and the
galaxy has a large radial velocity component, the actual length of the X-ray
tail may be larger. The 3 - 8 keV image clearly reveals a bright nuclear source.
We examined \gab's X-ray spectral properties in two regions, one within 17.4$''$
of the nucleus (note the 90\% encircled energy radius of the point-spread function
(PSF) at 1.49 keV is 14$''$ there)
and the other one excluding the nuclear region. The first region has 460$\pm$30
net counts in the 0.5 - 10 keV band, while the second region (the tail region)
has 253$\pm$29 net counts in the 0.5 - 2 keV band. The spectra are shown in Figure 14 and
the spectral fits are shown in Table 4. \gab\ hosts a Seyfert2-like nucleus.
The limited statistics on the soft X-rays, combined with the poor angular resolution
in the source position, prevent any definite conclusion on the gas properties.
A single-$kT$ fit implies a surprisingly hot tail with $kT \sim$ 2 keV. The gas
temperature can be reduced to $\sim$ 1.1 keV if a power-law component is included.
However, the power-law component has to be very strong (Table 4), given the small area of
the tail region (0.81 arcmin$^{2}$, compared with the case of \ga).
Like \ga's tail, the abundance of \gab's gas cannot be reliably constrained.
The X-ray gas mass of the tail can also be estimated.
We assume the tail as a cylinder with a length of 40 kpc and a radius of
2.5 kpc. Using the result from the one-$kT$ fit (the first row of Table 4), the total
gas mass is 4.2$\times10^{8}$ $f^{1/2}$ M$_{\odot}$, where $f$ is the filling
factor of the soft X-ray emitting gas. This may be an over-estimate as the gas
abundance is too low. In fact, the average electron density is 0.0188 $f^{-1/2}$ cm$^{-3}$.
The surrounding ICM has a temperature of $\sim$ 6.3 keV, similar to that around \ga.
The ICM electron density at the projected position is $\sim 1.9\times10^{-3}$ cm$^{-3}$
(B$\ddot{\rm o}$hringer et al. 1996). Thus, the X-ray tail would be $\sim$ 3 times
over-pressured. With the addition of a power-law component and a fixed abundance at
1 solar (the third fit in Table 4), the ISM electron density and the total gas mass
reduce to 0.0055 $f^{-1/2}$ cm$^{-3}$ and 1.2$\times10^{8}$ $f^{1/2}$ M$_{\odot}$
respectively. The last spectral fit may largely under-estimate the contribution from
the gas (see the difference of luminosities in Table 4). Thus,
we take an average mass estimate of $\sim 2\times10^{8}$ $f^{1/2}$ M$_{\odot}$.

The net H$\alpha$ image of \gab\ also shows an H$\alpha$ tail at the same position
as the X-ray tail, extending to at least 20 kpc from the nucleus (Figure 13).
A sharp H$\alpha$ edge is located only 1.1 kpc from the nucleus. H$\alpha$ absorption
is found upstream of the edge in the disk plane. We also searched for emission-line
objects in the field from the narrow-band data, hoping to find similar intracluster
HII regions as those downstream of \ga. However, no such emission-line regions have
been found in the sources detected in the H$\alpha$ band. The upper limit on the H$\alpha$ luminosity
(no intrinsic extinction) is 10$^{38}$ erg s$^{-1}$ and the upper limit on the
equivalent width of the H$\alpha$ line is 60 \AA.
Assuming a metallicity of 0.4 solar
and applying the Starburst99 model (Leitherer et al. 1999), this upper limit gives
a lower limit of 7 Myr for the age of any faded intracluster HII regions.
This may imply that \gab\ is in a more advanced stage of stripping and evolution than
\ga, which is consistent with the likely hotter X-ray tail of \gab\ and \gab's redder
optical color.
\gab\ is a more massive galaxy than \ga\ and may be undergoing edge-on stripping,
which indeed takes longer than the face-on stripping (e.g., RB08).

\section{Discussion}

\subsection{X-ray Tails of Late-type Galaxies and Mixing}

While the paucity of luminous X-ray tails of late-type galaxies in nearby clusters
has been known (e.g., Sun et al. 2007a), the discovery of \ga's double X-ray tails
is a surprise.
M86 also has double X-ray tails and Randall et al. (2008) interpreted it as the
consequence of the aspherical potential of M86 and M86's orbit. M86 is
an early-type galaxy so its potential structure is different from \ga's and
its tail is only composed of the stripped hot ISM. M86's tails are in fact connected to the
``plume'' on the north of the galaxy and are located at larger distance from the
galaxy, instead of coming from the galaxy directly like \ga's. Moreover,
a spectacular H$\alpha$ complex connecting M86 and the nearby disturbed spiral
NGC~4438 was recently discovered by Kenney et al. (2008). Some H$\alpha$ emission
also extends to the position of M86's tail. Thus, tidal interaction may
also be important for M86's X-ray tail. There also appears to be a split feature
that resembles a double tail in NGC~4438's \chandra\ image (Machacek et al. 2004;
Randall et al. 2008). However, one part of the split feature is almost along the
distorted disk plane and tidal interaction should be important in this case
(Machacek et al. 2004; Kenney et al. 2008). NGC~4438 is $\sim$ 3 times more
luminous than \ga\ in the $K_{\rm s}$ band. Its truncated halo should be larger than
\ga's ($\sim$ 15 kpc in radius, S07), also because the Virgo cluster is
less massive than A3627. The features in NGC~4438 only extend to $\sim$ 11 kpc
from the nucleus, clearly within its halo. The other late-type galaxy in the proximity,
NGC~4388, has many H$\alpha$ filaments and a spectacular HI tail (Yoshida et al. 2002;
Oosterloo \& van Gorkom 2005). A faint X-ray extension can be traced to
$\sim$ 14 kpc from the nucleus in the direction of the HI tail, from the
\chandra\ and \xmm\ data. There is also an X-ray extension to east in the
disk plane, which creates the appearance of double tails (Randall et al. 2008).
NGC~4388 is $\sim$ 1.5 times more luminous than \ga\ in the $K_{\rm s}$
band so similar to NGC~4438, its X-ray tail is still within its halo.
Thus, we do not consider the one-sided X-ray features in NGC~4438 and NGC~4388
to be X-ray double tails based upon the current data. They are clearly much
fainter than \ga's tails, especially given \ga's much larger distance
($\sim$ 4.3 times) and the much higher absorption to the direction of \ga.
We can also define strong X-ray tails as one-sided features outside of the
tidal truncation radius of the galactic halo (like the tails of \ga\ and \gab),
in order to distinguish them from the weak X-ray tails of NGC~4438 and NGC~4388.

The observed X-ray emission of the tails is likely from the interfaces of the hot
ICM and the cold stripped ISM, reflecting the multi-phase nature of mixing.
Although the detailed mixing process that determines the emergent X-ray spectra
is poorly understood, the observed tail temperature may be related to the
mass-weighted temperature that is usually adopted in simulations,
$T_{\rm mw} = (M_{\rm ISM} T_{\rm ISM} + M_{\rm ICM} T_{\rm ICM}) / (M_{\rm ISM} + M_{\rm ICM})
\approx T_{\rm ICM} / (1+X)$, where $X = M_{\rm ISM} / M_{\rm ICM}$ and we assume
that $T_{\rm ISM} \ll T_{\rm ICM}$. For the X-ray
tails of \ga\ and \gab, $X = 2 - 6.5$ if we assume $T_{\rm mw}$ is the observed
spectroscopic temperature. However, it is easy to find out that the mass-weighted
temperature is always larger than the spectroscopic temperature derived by fitting
the observed spectra with a single-$kT$ model ($T_{\rm spe}$).
We examined this with the CEMEKL and APEC models. For the CEMEKL model,
$d EM(T) \propto (T/T_{\rm max})^{\alpha-1} d T$.
Assuming an isobaric condition and for $\alpha >$ -1,

\begin{equation}
T_{\rm mw} = \frac{\int T dm(T)}{\int dm(T)} = \frac{1+\alpha}{2+\alpha} T_{\rm max}
\end{equation}

For $\alpha >$ -0.5, 1/3 $T_{\rm max} < T_{\rm mw} < T_{\rm max}$. It is also reasonable
to assume that $T_{\rm max} \leq T_{\rm ICM}$. However, with a series of XSPEC simulations,
we found that $T_{\rm spe} < T_{\rm mw}$. For example, $T_{\rm spe}$ = 0.5 - 1 keV when
-0.5 $< \alpha <$ 0.5, while $T_{\rm mw}$ = 2 - 3.6 keV for $T_{\rm max}$ = 6 keV.
This is not surprising as $T_{\rm spe}$ is mainly determined by the centroid of
the iron-L hump at low temperatures, which biases $T_{\rm spe}$ low. As long as
there are significant emission components at $kT$ = 0.4 - 2 keV, the iron-L hump
will be strong to make $T_{\rm spe}$
deviated from $T_{\rm mw}$. On the other hand, the parameter $\alpha$ in the
CEMEKL model may reflect the age of the X-ray tail. Our simulations indeed show that
$T_{\rm spe}$ increases monotonously with $\alpha$. From the CEMEKL fits to the spectra
of \ga\ and \gab, \gab's stripping event may indeed be older.
The above simple estimate implies that $T_{\rm spe}$ is a certain
fraction of the ICM temperature. The luminosity of the X-ray tail should also increase
with the magnitude of the ambient ICM pressure, under an isobaric condition.
Thus, the X-ray tails of late-type galaxies in groups and the outskirts of poor clusters
may have temperatures (e.g., $<$ 0.4 keV) and densities too low to be bright
in the \chandra\ and \xmm\ energy bands. This helps to explain the paucity of strong
X-ray tails of late-type galaxies in nearby clusters (e.g., Sun et al. 2007a).
Other factors include the generally high cluster background and the small number of
late-type galaxies in the \chandra\ field of nearby clusters as the covered cluster
volume is small.

With the above scenario, the X-ray tail of a late-type galaxy is simply the
manifestation of its cold ISM tail embedded in the hot ICM. Optical line emission
is also expected and is indeed detected, e.g., the H$\alpha$ tails of \ga\ and
\gab\ (S07 and Section 6). 
HI observations of A3627 were made with Australia Telescope Compact Array (ATCA)
in 1996 and no HI emission was detected from \ga\ or \gab\ (Vollmer et al. 2001a).
The 3$\sigma$ detection limit is $\sim$ 3 mJy/beam in one velocity channel with a
beamsize of 30$''$ (Vollmer et al. 2001a). Adjusting to the cluster distance used in
this work and assuming a linewidth of 150 km/s (Vollmer et al. 2001a), the corresponding
HI gas mass limit is
$\sim 5\times10^{8}$ M$_{\odot}$ per beam. While this limit is not weak for the
remaining HI gas in the disks of \ga\ and \gab\ (roughly within one beam), the X-ray
tails of \ga\ and \gab\ span over 7 - 15 times the beam size so a lot of diffuse HI
gas can remain undetected downstream of the galaxy. In fact, with the sensitivity
of the available ATCA data (an HI column density limit of 2$\times10^{20}$ cm$^{-2}$),
four of the seven HI tails in the Virgo cluster found by Chung et al. (2007) and the
long HI tail of NGC~4388 (Oosterloo \& van Gorkom 2005) would not be detected.
Clearly deeper HI observations are
required, although the nearby bright radio source PKS~1610-60 (43 Jy at the 1.4 GHz)
makes the task difficult (8$'$ - 14$'$ from \ga\ and \gab).
Both \ga\ and \gab\ are expected to host a significant amount of atomic and
molecular gas initially. From Blanton \& Moustakas (2009), the total neutral plus
molecular gas fraction (relative to the stellar mass) is $\sim$ 70\% for \ga\ and
$\sim$ 30\% for \gab. Assuming a type of Sc and a blue-band diameter of 20 kpc, the
expected HI mass of \ga\ is $\sim 2.4\times10^{9}$ M$_{\odot}$, from the empirical
relation derived by Gavazzi et al. (2005). Similarly for \gab\ (a type of Sa and
a blue-band diameter of 30 kpc), the expected HI mass is
$\sim 3.9\times10^{9}$ M$_{\odot}$. Thus, \ga\ is expected to have
$\sim (3-5)\times10^{9}$ M$_{\odot}$ of cold ISM initially, while \gab\ is expected to
have $\sim (4-9)\times10^{9}$ M$_{\odot}$ initially. These numbers can be compared
with the X-ray gas mass in their X-ray tails (Table 2).
Sivanandam et al. (2010) detected the mid-IR emission from warm H$_{2}$ gas
($\sim$ 150 K) in the galaxy and the first 20 kpc of the main tail, with a total
mass of $\sim 2.5\times10^{7}$ M$_{\odot}$. The H$_{2}$ lines that IRS detects
are rotationally excited lines that are generated very efficiently. The warm gas
they probe is likely a small fraction of the total H$_{2}$ gas, if the bulk of the
molecular gas is still cold. Deep HI and CO observations are required to recover
the bulk of the cold gas in \ga, \gab\ and their wakes.

The same question can also be asked to other cluster late-type galaxies with signs of
stripping. Does a stripped ISM tail show up in HI, H$\alpha$ and X-rays
simultaneously? We attempt to summarize the known examples of stripped tails
for cluster late-type galaxies. A Venn diagram is plotted in Figure 15. While a lot
of stripped tails are known for cluster late-type galaxies, there is currently little
known overlap between different bands, especially between X-ray tails and HI tails.
This may be due to the lack of sufficient data, either in X-rays (e.g.,
for the Virgo galaxies with HI tails, Chung et al. 2007) or in HI (e.g.,
the two galaxies in this
paper). We are pursuing \xmm\ data for the Chung et al. (2007) galaxies and
deeper ATCA data for the A3627 galaxies to better address this question.
On the other hand, there are various reasons that A one-to-one correlation may
not exist in every case, even with better data.
Depending on its age and the surrounding ICM, an ISM tail may emit predominantly
in the HI band or the X-ray band. Soft X-ray emission of the tail may only be
bright in high-pressure environment.
We also understand little about the details of mixing. It may be related
to the mean free path of particles and the coherence length of the magnetic field,
which would introduce a radial dependence on the efficiency of mixing.
Nevertheless, we first need better data to update the Venn diagram shown in
Figure 15.

Deep X-ray data alone, like \ga's, can reveal some thermal history of the
stripped ISM. One surprise from \ga's X-ray data is the constancy of
the spectroscopic temperatures along both tails (Figure 7). Indeed extra care is
required to interpret this spectroscopic temperature as it is mainly determined
by the centroid of the combined iron-L humps from multi-$T$ gas. Nevertheless,
the possible spectral difference between \ga's tails and \gab's tail is
intriguing. This difference, if confirmed, combined with the temperature
constancy of \ga's tails, puts a constraint on the heating timescale of the
stripped cold ISM.
Clearly the knowledge of the HI/CO gas distribution
in these two galaxies will help to tighten the constraint, although it is
certain that this timescale depends on the ICM temperature.

\subsection{Comparison with Simulations}

After the initial analytic work by Gunn \& Gott (1972), many simulations have
been run on the ram-pressure stripping of late-type galaxies in clusters
(e.g., Abadi et al. 1999; Schulz \& Struck 2001; Vollmer et al. 2001b; RB08;
Kapferer et al. 2009; Tonnesen \& Bryan 2009
and references in those papers). Recent efforts surpass the early ones in many
respects, e.g., resolution, time steps, realistic treatment of the galaxy orbit
and some important baryon physics (e.g., cooling and star formation). We mainly compare
our results with recent simulations of three groups, the Innsbruck group with the
GADGET-2/SPH code (Kronberger et al. 2008; Kapferer et al. 2009), the Bremen group
with the FLASH/AMR code (Roediger et al. 2006; RB08)
and the Columbia group with the {\em Enzo}/AMR code (Tonnesen \& Bryan 2009).
The GADGET-2 simulations include recipes for cooling, star formation and stellar
feedback. The model galaxies run through a wind tunnel with a constant pressure.
The FLASH simulations are non-radiative and model the flight of a disk galaxy through a galaxy cluster
with realistic ICM and potential distributions. The {\em Enzo} simulations include
cooling but not feedback (Tonnesen \& Bryan 2009). To mimic effects of some heating
processes not in simulations, they impose a minimum temperature floor.
The model galaxy is also subject to a constant ram pressure.
The inclusion of cooling is important, as it indeed happens as revealed from our data
(the intracluster HII regions and star clusters) and it has profound effects in
simulations (Kapferer et al. 2009; Tonnesen \& Bryan 2009). Cooling also naturally
adjusts the ISM distribution on the disk plane, creating multi-phase ISM with enhancement
and holes (Kapferer et al. 2009; Tonnesen \& Bryan 2009), which is not present
in the non-radiative runs. However, cooling needs
to be offsetted by stellar feedback. Unknown efficiencies of transport
processes (viscosity and heat conduction) because of unclear strength and configuration
of magnetic field, plus the other uncertainties in the prescriptions of star formation and
stellar feedback, make the tasks difficult (e.g., Tonnesen \& Bryan 2009).
We emphasize that comparison at the current stage is not straightforward as 
galaxies in these simulations are always more massive than \ga\ and high pressure
environment was not explored in all simulations.
We also caution that it is best to compare the observational data with the mock data
generated from simulations. Kapferer et al. (2009) and Tonnesen \& Bryan (2009)
attempted to create X-ray mock data but more work is required, e.g., folding with the
actual \chandra\ response, projection on the ICM background and the use of spectroscopic
temperature rather than mass-weighted temperature (Section 7.1).

Tails generally appear longer and wider in simulations than in observations.
The simulations by Kapferer et al. (2009) show that the X-ray tails can be
traced to $\sim$ 300 kpc from a galaxy that is about 4 times more massive
than \ga, with an ambient pressure similar to \ga's. The simulated X-ray
tails are also very clumpy, with widths generally increasing with the distance
from the galaxy. The FLASH simulations also produce longer tails than observed
in this work, but are difficult to compare with the X-ray data directly. As
RB08 pointed out, the observed tail length depends on the mass-loss per
orbital length, which is the highest in poor clusters as galaxies in rich
clusters move too fast and the ram pressure in groups is not high enough.
The FLASH simulations also show tail flaring (i.e. increasing width with
distance from the galaxy downstream) in all cases.
The velocity perpendicular to the direction of the ram pressure
comes from the random motion of the ICM, the disk rotation and turbulence
in the wake. For a rotation velocity of 100 km/s over 50 Myr, the offset is
5 kpc, which is not small in \ga's case as its tails are narrow.
The Tonnesen \& Bryan (2009) simulations produce long but narrower tails.
The tails are also very clumpy. None of these simulations can reproduce the
double tails of \ga. The simulated tails are generally too clumpy with strong
turbulence. This simple comparison may point to a higher ICM viscosity than
present in the simulations. The Reynold number is $\sim 3{\cal M}(L/\lambda)$, where
${\cal M}$ is the Mach number, $L$ is the size of the remaining galactic gas disk
and $\lambda$ is the mean free path of particles in the ICM. For unmagnetized gas,
$\lambda \sim$ 10 kpc around \ga. The radius of the remaining H$\alpha$ disk 
and the galactic X-ray emission is only $\sim$ 1.7 kpc (S07). Thus, the Reynold
number is on the order of unity which implies a laminar flow. Of course the
presence of magnetic field would increase the Reynold number.
In the simulations by Kapferer et al. (2009), a bright spot in the X-ray
images is always found at distances of a few tens of kpc from the galaxy,
caused by compressional heating. This feature can be compared with the bright
spot in the main tail (Figure 1 and 5). However, the location of this feature
is generally too far from the galaxy in simulations with high ambient pressure.
Such a bright spot is also not observed in the secondary tail.

The most obvious question about \ga's X-ray tails is why are there double tails.
The simplest explanation is to involve another dwarf galaxy forming a pair with
\ga. This was in fact simulated by Kapferer et al. (2008), showing double gaseous
tails, stellar tidal tails and a stellar bridge between two galaxies.
The X-ray gas mass of the secondary tail is 40\% - 50\% of the mass of the main
tail, assuming the same filling factors of the X-ray gas. If we simply take
the ratio of their X-ray gas masses as the ratio of their ISM gas masses,
the stellar mass of this assumed dwarf that is responsible for the secondary
tail should be $\sim$ 30\% of \ga's stellar mass, from the ISM mass - stellar
mass relation summarized by Blanton \& Moustakas (2009).
Such a galaxy should be easily detected, at least through the tidal tails and bridge
it causes even if it hides behind \ga. However, the SOAR images do not show
the presence of another galaxy near \ga. There are also no tidal features
upstream of the galaxy. There are some continuum features downstream of
the galaxy as shown in S07. However, all of them are around confirmed HII
regions and have blue colors. Our new \hst\ data only reinforce the conclusion
from the SOAR images. The central region of the galaxy is very dusty but the
galaxy appears regular without significant tidal features in the \hst\ $I$ band
(Paper IV, in preparation).
Thus, we conclude that both X-ray tails originate from \ga.

The discovery of two separated stripping tails from one galaxy presents
a challenge to our understanding of stripping. In simulations (e.g.,
RB08; Kapferer et al. 2009; Tonnesen \& Bryan 2009), cold ISM is removed from
the sides of the disk or through holes and is quickly mixed to form one
clumpy tail because of disk rotation, momentum transferred from the ICM
and turbulence in the wake, especially if projected on the plane of sky.
Perhaps we should begin to consider more realistic ISM distributions in the galaxy.
Optical images clearly show the presence of two thick spiral arms, one to
the north and the other one to the south (S07; better shown in the
\hst\ images). Maybe the two X-ray tails correspond to stripping of
molecular gas concentrated around these two spiral arms.
CO images of nearby face-on spiral galaxies indeed show that molecular gas
distribution follows major spiral arms in many cases (e.g., Kuno et al. 2007).
This is also generally true for HI gas (e.g., Adler \& Westpfahl 1996).
For \ga, we present a simple cartoon in Figure 16.
The north arm is apparently the trailing arm where stripping is easier so
a brighter X-ray tail is produced. The south arm is the leading arm so gas
removal there is more difficult as most gas would be pushed inwards to deeper
galactic potential first.
This also explains the fact that more intracluster HII regions (at $>$ 15 kpc
from the nucleus) are found around the main tail while the
numbers of HII regions in the halo are similar at two sides. 
The curvature of the secondary tail may come from the galactic
rotation, which is high enough to cause the observed offset. The main tail
may also be curved in 3D but appears straight in projection.
The rather regular wake structure may imply significant viscosity.
On the other hand, draping of the ICM magnetic field (e.g., Dursi \& Pfrommer 2008)
can also inhibit the motion of the stripped ISM perpendicular to the direction
of ram pressure.

In summary, it remains to be seen whether simulations can reproduce the
double X-ray tails of \ga\ as observed, and the gas properties in other
bands (H$\alpha$ and H$_{2}$ at least). We also emphasize that the
comparison between data and simulations is not trivial and straightforward.
Observationally we need
deeper data at multiple wavelengths. For simulations, mock data need to
be produced with sensitivities matching the actual observations.
Uncertainties in the viscosity, heat conduction, cooling and stellar feedback
should be explored with different runs. Eventually we need to compare the
multi-wavelength data (HI, CO, IR, optical and X-rays) with the mock data
in the same band. Besides the simple comparison of morphology, a lot of details
of the tail properties (e.g., temperature distribution, filling factors
of the gas in different phases and energy transfer between gas in
different phases) can be better examined.

\subsection{Formation of Young Stars and X-ray Binaries in the Stripped ISM}

The new observations of \ga, for the first time, unambiguously confirm the active star
formation in the cold ISM stripped by the ICM ram pressure. As emphasized in S07,
\ga's immediate surroundings are devoid of bright galaxies. There are only two
galaxies that are within 4 mag of \ga\ in the $I$ band, within 100 kpc from
the end of \ga's X-ray tail (G1 and G2 in Figure 1 of S07). Their redshifts are
unknown and they are 1.9 - 2.7 mag fainter than \ga\ in the $I$ band. There is also no
such galaxy within 70 kpc from the nucleus of \ga. Both G1 and G2 are
undisturbed and there are no tidal features between \ga\ and G1 (or G2).
Our \hst\ data do not reveal another dwarf galaxy near \ga\ (Paper IV,
in preparation). The velocity map of the HII regions also shows no extra
component, besides \ga's rotation pattern. It is also known that the cluster
potential is not the main factor for stripping (S07; Sivanandam et al. 2010),
although it indeed affects the size of \ga's dark matter halo.
Thus, we conclude that ICM ram pressure is responsible for the stripping of the
ISM. In our case, the ambient pressure is even high enough to remove some molecular
gas from the galaxy. The HI gas has typical pressures of 
$P/k_{\rm B} \sim 10^{3} - 10^{4}$ K cm$^{-3}$, while the pressure in the core of a
starless giant molecular cloud can be higher than $10^{6}$ K cm$^{-3}$. The ambient
total thermal pressure (electron + ions) is 1.3$\times10^{5}$ K cm$^{-3}$ for
$n_{\rm e}$ = 10$^{-3}$ cm$^{-3}$
and $kT$=6 keV, while the ram pressure is 3.2$\times10^{5}$ K cm$^{-3}$ for $n_{\rm e}$ =
10$^{-3}$ cm$^{-3}$ and $v_{\rm gal}$ = 1500 km/s.

After removing the cold ISM from the galaxy, the key for the intracluster star formation
is that the stripped ISM must be able to cool, likely aided by shocks and the absence
of the UV radiation field that is strong on the disk plane.
Intracluster star formation in the stripped ISM was first suggested
in simulations (Schulz \& Struck 2001; Vollmer et al. 2001b). Schulz \& Struck (2001)
even suggested formation of dwarf galaxies in this mode. In the new GADGET-2/SPH
simulations by Kronberger et al. (2008) and Kapferer et al. (2009), new stars are
formed in the stripped ISM, forming $\sim$ 1 kpc structures with masses of up to
10$^{7}$ M$_{\odot}$ individually to hundreds of kpc downstream of the galaxy.
However, all simulations suffer from uncertainties in transport processes
like viscosity and heat conduction, which are important for determining the heating
rate.

More evidence of intracluster star formation, either HII regions or blue star
clusters, has been revealed recently (e.g., Lee et al. 2000; Gerhard et al. 2002;
Cortese et al. 2004; Cortese et al. 2007; Yoshida et al. 2008). They are only at
one side of the galaxy. In some cases, tidal interaction may be important (e.g.,
Lee et al. 2000). But they all lack unambiguous associations with gaseous tails.
We especially highlight the growing number of examples of one-sided blue stellar filaments
and star clusters behind cluster galaxies, which includes two galaxies in
A2667 and A1689 at $z \sim$ 0.2 (Cortese et al. 2007) and a dwarf galaxy RB~199
in the Coma cluster (Yoshida et al. 2008). The connection between the first two
galaxies at $z \sim 0.2$ with \ga\ was discussed in S07.
In the case of RB~199, Yoshida et al. (2008) found narrow blue filaments, knots, H$\alpha$
filaments and clouds extending to 80 kpc from the galaxy in one side. It is
a rather remarkable case as the galaxy is much fainter than all
previous galaxies, with a stellar mass of $\sim 8 \times 10^{8}$ M$_{\odot}$.
The stellar mass of the downstream structures is $\sim 10^{8}$ M$_{\odot}$,
which is $\sim$ 10\% of the galactic mass.
Yoshida et al. (2008) also favored a mechanism similar to what we suggested for \ga\
in S07. These three examples may imply a rather common evolution stage for cluster
late-type galaxies. As the stage with bright blue stellar trails is short ($\leq$ 200 Myr),
more similar examples are expected only in large surveys.

If we assume that intracluster star formation happens in the stripped ISM of
every cluster spiral, can this mode contribute a significant fraction of
the intracluster light? The answer depends on the mass fraction of intracluster
stars formed out of the stripped ISM (or the star formation efficiency).
S07 gave a rough estimate of $\sim$ 1\%,
which may only be a lower estimate as blue stellar filaments and clusters
(not shining in H$\alpha$) are not counted. While a detailed account of
the total stellar mass downstream of \ga\ is the aim of Paper IV, here we take
1\% as a lower limit. If the efficiency of the intracluster star formation
is only 1\%, the intracluster light contributed from star formation in stripped ISM
is small, given the mass fraction of the intracluster light to the total
stellar light in clusters ($\sim$ 20\%, likely depending on the halo mass,
e.g., Krick \& Bernstein 2007). However, the case of RB~199 implies a fraction	
of $\sim$ 10\%, if the original ISM in RB~199 had a mass comparable to its
stellar mass (e.g., Blanton \& Moustakas 2009). 
In the simulations by Kronberger et al. (2008) and Kapferer et al. (2009),
7\% - 12\% of the initial ISM turns into new stars in the wake.
If an efficiency of 10\% is adopted, the intracluster light contributed by
ram pressure stripping is a significant fraction of the total
intracluster light. However, this efficiency may not be too high given the
star formation efficiency in galaxies. In galactic disks, $\sim$ 1\% of gas
is converted to stars per free-fall time (e.g., Leroy et al. 2008).
In 100 Myr (a typical orbital time in the disks), $\sim$ 6\% of gas is converted
to stars (e.g., Kennicutt 1998; Leroy et al. 2008). We note that 100 Myr is
also about the orbital time of the residual H$\alpha$ disk ($\sim$ 1.7 kpc radius with an
orbital velocity of $\sim$ 100 km/s) and close to the age of the X-ray tails.
Clearly, a lot of work is required to constrain the intracluster star formation
efficiency. The efficiency may also change with the mass of the host galaxy as
more stars are formed inside the halo of more massive galaxies.
In principle, the ratio of the intracluster SNIa to the other types can
also inform us the significance of intracluster star formation.

As active star formation is confirmed downstream of the galaxy,
the discovery of ULXs is not a surprise.
As we estimated in Section 5.2, the total luminosity of ULXs can be
consistent with the expectation from the intracluster star formation rate.
The space density of X-ray point sources also traces the
strength of the star formation, stronger in the halo and weaker
around the unbound tail.
A similar system may be the high-velocity system of NGC~1275.
Gonzalez-Martin et al. (2006) found a concentration of eight ULXs
($L_{\rm 0.5-7 keV} > 3\times10^{39}$ erg s$^{-1}$) around the high-velocity system.
They may be associated with the active star formation in and downstream of the
high-velocity system, similar to what is happening in \ga.

\section{Summary and Future Work}

We present the discovery of spectacular X-ray double tails behind \ga\
in the closest rich cluster A3627 from our 140 ks \chandra\ observation.
\gem\ spectroscopic data of HII regions downstream of \ga\ are also discussed.
We also present the discovery of a likely heated X-ray tail associated with
\gab, also in A3627. The main results of this paper are:

1) Besides the known main X-ray tail of \ga, a fainter secondary tail is
revealed with the deep \chandra\ observation, as well as some substructures
(Section 3 and Figure 1 and 2). The double tails of \ga\ make it morphologically
resemble radio head-tail galaxies. The whole tail system exhibits intriguing
regularity (almost constant tail widths, constant spectroscopic temperatures
and well separated tails not twisted together) and some complexity
(substructures). We consider that the soft X-ray emission of the tails come
from the mixing of the stripped cold ISM with the hot ICM.
We also conclude that the two tails are detached and both of them
originate from \ga, stripped by the ICM ram pressure. One possible scenario
(Section 7.2) involves the two major spiral arms of \ga\ that is visible from the
optical data, dense ISM (especially molecular gas) concentration around two arms,
the orientation of the arms to the ram pressure and a weak level of turbulence
downstream of the galaxy (or a higher viscosity than that in the current simulations).

2) The \chandra\ spectra reveal the multi-phase nature of the X-ray tails,
requiring more than one thermal component to obtain good fits and reasonable
abundance (e.g., 1 solar). However, the X-ray tails are always over-pressured
to the ICM with the current simple models (Section 3.2).
We also examined the spectroscopic temperature distribution in
the tails with a single-$kT$ model. \ga's double tails are surprisingly
isothermal ($\sim$ 0.8 keV), embedded in $\sim$ 6 keV ICM (Figure 7). 

3) \ga's X-ray tails also present a challenge to the current simulations
of stripping of cluster spirals (Section 7.2). Nevertheless, we emphasize
that the comparison is not trivial and straightforward. Mock data from
simulations need to be generated with sensitivities matching the actual
observations. Eventually we need multi-wavelength data (e.g.,
HI, CO, IR, optical and X-rays) in order to examine the interplay and energy
transfer between gas in different phases (Section 7.1 and Figure 15).

4) \gem\ spectra were obtained for 33 emission-line regions identified by S07
(Section 5.1). All of them are confirmed as HII regions downstream of \ga\
(Section 5.1 and Figure 9-11). Their velocity map shows the imprint of
\ga's rotation pattern (Figure 10). For the first time, we unambiguously
know that active star formation can happen in the cold ISM stripped by ICM
ram pressure. We also summarized the current known cases of intracluster
star formation, especially ones with one-sided blue star clusters and stellar
trails (Section 7.2).
The intracluster  star-formation efficiency is expected to be higher than
1\%. The contribution to the intracluster light from this mode can be high
if the star-formation efficiency is as high as 10\% (Section 7.3).

5) This deep observation also reveals many X-ray point sources (Section 5.2). A significant
source excess is observed immediately downstream of \ga. Four of the X-ray point
sources are within 0.65 kpc from spectroscopically confirmed HII regions.
We consider them strong candidates for intracluster ULXs with
$L_{\rm 0.3-10 keV}=0.92-6.6\times10^{39}$ erg s$^{-1}$. Other candidates
(at least two) have $L_{\rm 0.3-10 keV}$ up to 2.5$\times10^{40}$ erg s$^{-1}$.
We also conclude that the observed total X-ray luminosity of the intracluster
ULXs is consistent with the expectation from the observed total star formation rate
of HII regions.

6) A 40 kpc X-ray tail was detected behind another late-type galaxy in A3627,
\gab\ (Section 6). Signatures of stripping were also found in the H$\alpha$ data, with a sharp
H$\alpha$ edge and an H$\alpha$ tail extending to at least 20 kpc from
the nucleus. No HII regions are found downstream of \gab. Its X-ray tail
is also surprisingly hot ($kT \sim$ 2 keV) but the uncertainty is big.
We suggest that \gab's stripping event is at a more advanced stage than \ga's.

Why are there two bright X-ray tails of late-type galaxies in A3627? Besides
coincidence, the high ambient pressure (ram pressure + thermal pressure) in A3627
should also play a role, making soft X-ray tails denser and more luminous.
This would explain the absence of bright X-ray tails of late-type galaxies
in the Virgo cluster (see Section 7.1).
A high-pressure environment also help intracluster star formation, by moving
high-density molecular gas into the intracluster space.

This paper mainly focuses on the \chandra\ data of \ga.
In a future paper, we will examine the intracluster star formation
downstream of \ga\ in more detail, from the \hst\ and \gem\ data.
HI and CO data will also be pursued.
With a more complete set of multi-wavelength data, the nature of the X-ray tails
and the intracluster star-formation could be better understood. \ga\ provides the
best opportunity to simultaneously study stripping and the related important
questions across the electromagnetic spectrum.

\acknowledgments

M. S. is grateful to Tom Matheson for his helps in the planning and
the data analysis of the \gem\ observations.
M. S. thanks Suresh Sivanandam for the preprint of the \spi\ results and discussions.
We thank Hans B$\ddot{\rm o}$hringer, Wolfgang Kapferer and Bernd Vollmer for discussions.
We thank Keith Arnaud for the clarification on the CEMEKL model.
We present data obtained at the Gemini Observatory, which is operated by the
Association of Universities for Research in Astronomy, Inc., under a cooperative agreement
with the NSF on behalf of the Gemini partnership.
We also present data obtained with the Southern Observatory for
Astrophysical Research, which is a joint project of the Brazilian
Ministry of Science, the National Optical Observatories, the University
of North Carolina and Michigan State University. We made use of the
NASA/IPAC Extragalactic Database and the Hyperleda database.
The financial support for this work was provided by the NASA grant GO7-8081A,
the NASA grant GO8-9083X and the \hst\ grant HST-GO-11683.01-A.
MD and GMV acknowledge support from a NASA LTSA grant NASA NNG-05GD82G.
ER acknowleges funding by the DFG Priority Programme 1177 `Galaxy Evolution'.

{\it Facilities:} \facility{CXO (ACIS)}, \facility{Gemini:South (GMOS)}, \facility{SOAR (SOI)}.

\begin{table}
\begin{center}
{\small
\caption{Spectral fits for the global tail spectrum of \ga$^{a}$}
\begin{tabular}{lcc} \hline \hline
 Model$^{b}$ & Parameters$^{c}$ & C-statistic$^{d}$ (d.o.f) \\ \hline

 APEC & $kT$=0.805$\pm$0.033, $Z$=0.06$\pm$0.02 & 40.3 (46) \\
 APEC & $kT$=0.780$\pm$0.020, $Z$=(0.5) & 105.1 (47) \\
 APEC+APEC & $kT_{1}$=0.719$^{+0.048}_{-0.064}$, $Z_{1}$=0.20$^{+0.51}_{-0.10}$, $kT_{2}$=2.0$^{+1.6}_{-0.4}$, $Z_{2}$=(0.20), NORM$_{1}$/NORM$_{2} \sim 1.4$ & 30.1 (44) \\
 APEC+APEC & $kT_{1}$=0.668$^{+0.059}_{-0.039}$, $Z_{1}$=(1.0), $kT_{2}$=1.7$^{+0.3}_{-0.2}$, $Z_{2}$=(0.20), NORM$_{1}$/NORM$_{2} \sim 0.2$ & 31.0 (45) \\
 APEC+PL & $kT$=0.761$^{+0.037}_{-0.033}$, $Z$=0.10$^{+0.04}_{-0.03}$, $\Gamma$=(1.7), $L_{\rm PL, 0.3-10 keV}=4.5\times10^{40}$ & 32.5 (45) \\
 APEC+PL & $kT$=0.754$^{+0.038}_{-0.031}$, $Z$=(1.0), $\Gamma$=2.36$\pm$0.16, $L_{\rm PL, 0.3-10 keV}=1.0\times10^{41}$ & 33.5 (46) \\
 CEMEKL & $\alpha$=0.47$^{+0.46}_{-0.21}$, $kT_{\rm max}$=2.8$^{+1.9}_{-1.2}$, $Z$=0.35$^{+0.32}_{-0.10}$ & 34.7 (45) \\
 CEMEKL & $\alpha$=0.37$\pm$0.15, $kT_{\rm max}$=6.0$^{+3.0}_{-1.6}$, $Z$=(1.0) & 37.0 (46) \\

\hline \hline
\end{tabular}
\vspace{-1cm}
\tablenotetext{a}{The spectrum (shown in Figure 8) is extracted from the whole tail region
(green regions in Figure 7 excluding the head region). Point sources are excluded.}
\tablenotetext{b}{The Galactic absorption component (1.73$\times10^{21}$ cm$^{-2}$) is included in all cases. CEMEKL is a multi-temperature plasma emission model built from the MEKAL 
code. Emission measure follows: $EM(T) \propto (T/T_{\rm max})^{\alpha}$.}
\tablenotetext{c}{The unit of $kT$ is keV and the unit of $Z$ is solar. The unit of luminosity
is erg s$^{-1}$. Parameters in parentheses are fixed. We consider the ``APEC+APEC'' model
and the ``CEMEKL'' model are better physical descriptions of the data than the other models tried.}
\tablenotetext{d}{C statistic does not provide a goodness-of-fit
measure but the XSPEC version of the C statistic is defined to approach $\chi^{2}$
in the case of large number of counts. This number is only quoted as a reference for
comparison of different models.}
}
\end{center}
\end{table}

\begin{table}
\begin{center}
{\small
\caption{The properties of \ga\ and \gab}
\begin{tabular}{lcc} \hline \hline
Property & \ga\ & \gab\ \\ \hline

 Heliocentric velocity (km/s)$^{a}$ & 4660 (-211) & 5743 (+872) \\
 log(L$_{\rm Ks}$/L$_{\odot}$)$^{b}$ & 10.42 (10.18) & 11.11 (11.20) \\
 $B-K_{\rm s}$ (mag)$^{c}$ & 2.76 (2.15) & 3.18 (3.40) \\
 $M_{*}$ (10$^{9}$ M$_{\odot}$)$^{d}$ & 5 - 8 & 32 - 39 \\
 $l \times w$ (kpc $\times$ kpc)$^{e}$ & $\sim 80 \times$ 8, 80 $\times$ 7 & $\sim 40 \times$ 5 \\
 $kT$ (keV)$^{f}$ &   0.80$\pm$0.04 & 1.98$^{+0.96}_{-0.56}$ \\
 $f_{\rm 0.5-2 keV, thermal, obs}$ (erg cm$^{-2}$ s$^{-1}$) (tail)$^{f}$ & (8.3$\pm$0.4)$\times10^{-14}$ & (3.4$\pm$0.5)$\times10^{-14}$ \\
 $L_{\rm 0.5-2 keV, thermal}$ (erg s$^{-1}$) (tail)$^{f}$ & (8.3$\pm$0.4) $\times10^{40}$ & (3.2$\pm$0.5) $\times10^{40}$ \\
 $L_{\rm bol, thermal}$ (erg s$^{-1}$) (tail)$^{f}$ & (1.8$\pm$0.1) $\times10^{41}$ & (7.8$\pm$1.2) $\times10^{40}$ \\
 $L_{\rm 0.5-2 keV, thermal}$ (erg s$^{-1}$) (total)$^{f}$ & (1.10$\pm$0.05) $\times10^{41}$ & (5.5$\pm$0.9) $\times10^{40}$ \\
 $L_{\rm 0.3-10 keV, nucleus}$ (erg s$^{-1}$)$^{g}$ & $< 1.2\times10^{39}$ & (1.4$\pm$0.2) $\times10^{42}$ \\
 $n_{\rm e}$ ($f^{-1/2}$ cm$^{-3}$)$^{h}$ & 0.0073 - 0.019 & 0.0055 - 0.019 \\
 $M_{\rm gas}$ ($10^{9} f^{1/2}$ M$_{\odot}$)$^{h}$ & $\sim 1$ & $\sim 0.2$ \\

\hline \hline
\end{tabular}
\vspace{-1cm}
\tablenotetext{a}{The velocity values in parentheses are the difference from A3627's
system velocity, 4871 km/s (Woudt et al. 2008). \gab's velocity is from Woudt et al. (2004).} 
\tablenotetext{b}{The $K_{\rm s}$ band magnitudes are from Skelton et al. (2009), while the values in parentheses are from the Two Micro All Sky Survey (2MASS).}
\tablenotetext{c}{The intrinsic reddening is not corrected. The values in parentheses are for 2MASS $K_{\rm s}$ magnitudes.}
\tablenotetext{d}{The stellar mass of the galaxy, estimated from the $I$ band light (from S07)
and the $K_{\rm s}$ band light (from Skelton et al. 2009), with the method discussed in
Section 4.2 of S07. The estimates will be smaller if intrinsic reddening exists as galaxies will be bluer.}
\tablenotetext{e}{The approximate dimension of the tails}
\tablenotetext{f}{From the single-$kT$ fits. The listed fluxes are observed values for the thermal gas in the tail while the luminosities are unabsorbed values.}
\tablenotetext{g}{Assuming no intrinsic absorption for \ga's nucleus. If the intrinsic absorption is
10$^{23}$ cm$^{-2}$, the limit is $< 1.1\times10^{40}$ erg s$^{-1}$.}
\tablenotetext{h}{$f$ is the filling factor of the soft X-ray emitting gas.}
}
\end{center}
\end{table}

\begin{table}
\begin{center}
{\scriptsize
\caption{The properties of \chandra\ point sources around \ga\ and its tails}
\begin{tabular}{ccccccc} \hline \hline
Source \#$^{a}$ & RA$^{b}$ & Dec$^{b}$ & 0.5 - 7 keV counts$^{c}$ & $f_{\rm 0.3-10 keV}^{d}$ & $\Gamma$ & note \\
          & (J2000) & (J2000) & & (10$^{-15}$ erg s$^{-1}$ cm$^{-2}$) & & \\ \hline

1  & 16:13:30.43 & -60:45:18.2 & 17.3$\pm$5.1 & 1.6$\pm$0.5 & (1.7) & a bright stellar object \\
2  & 16:13:29.76 & -60:46:30.0 & 47.3$\pm$7.6 & 12.4$\pm$2.0 & 0.02$\pm$0.31 & a faint optical source \\
3  & 16:13:27.66 & -60:46:22.7 & 35.4$\pm$6.6 & 3.2$\pm$0.6 & (1.7) & a bright stellar object \\
4  & 16:13:27.75 & -60:45:34.4 & 17.5$\pm$5.0 & 1.6$\pm$0.5 & (1.7) & no optical counterpart \\
5  & 16:13:27.44 & -60:45:54.4 & 41.4$\pm$7.3 & 3.7$\pm$0.7 & (1.7) & no optical counterpart \\
6  & 16:13:26.03 & -60:45:38.3 & 13.9$\pm$4.6 & 1.3$\pm$0.4 & (1.7) & no optical counterpart \\
7  & 16:13:25.80 & -60:45:45.3 & 64.6$\pm$9.1 & 6.6$\pm$1.0 & 3.45$\pm$0.44 & a bright stellar object \\
8*  & 16:13:25.76 & -60:45:10.3 & 33.3$\pm$6.7 & 3.0$\pm$0.6 & (1.7) & between two HII regions, 1.3$''$ from them \\
9  & 16:13:24.43 & -60:45:45.2 & 115.1$\pm$11.6 & 9.8$\pm$1.0 & 1.87$\pm$0.17 & 4.0$''$ from an HII region \\
10* & 16:13:24.30 & -60:46:06.6 & 42.0$\pm$7.2 & 3.5$\pm$0.6 & 2.17$\pm$0.33 & 2.0$''$ from an HII region \\
11* & 16:13:24.29 & -60:45:37.2 & 116.5$\pm$11.7 & 11.4$\pm$1.2 & 1.54$\pm$0.19 & 0.8$''$ from an HII region \\
12 & 16:13:23.57 & -60:45:29.8 & 414.3$\pm$21.0 & 42.6$\pm$2.2 & 1.38$\pm$0.09 & 2.6$''$ from an HII region \\
13 & 16:13:18.28 & -60:43:55.8 & 32.6$\pm$6.6 & 3.0$\pm$0.6 & (1.7) & no optical counterpart \\
14 & 16:13:16.94 & -60:44:51.5 & 23.0$\pm$5.7 & 2.1$\pm$0.5 & (1.7) & a bright stellar object \\
15* & 16:13:16.76 & -60:44:18.1 & 16.8$\pm$4.8 & 1.6$\pm$0.5 & (1.7) & 1.0$''$ from an HII region \\
16 & 16:13:10.74 & -60:44:37.6 & 16.5$\pm$4.9 & 1.5$\pm$0.5 & (1.7) & a bright stellar object \\
17 & 16:13:10.19 & -60:45:04.0 & 32.7$\pm$6.5 & 3.0$\pm$0.6 & (1.7) & a bright stellar object \\
18 & 16:13:07.12 & -60:44:54.3 & 17.0$\pm$4.9 & 1.6$\pm$0.5 & (1.7) & no optical counterpart \\
19 & 16:13:06.34 & -60:43:38.0 & 48.4$\pm$7.9 & 4.9$\pm$0.8 & 1.38$\pm$0.32 & a faint optical source \\

\hline \hline
\end{tabular}
\vspace{-1cm}
\tablenotetext{a}{Sources with * are considered strong candidates of intracluster ULXs. Sources 9 and 12 are also candidates.}
\tablenotetext{b}{The position uncertainty is $\sim 0.4''$ from the examination of eight stars in the field.}
\tablenotetext{c}{Counts and errors are from ``WAVDETECT'' of CIAO.}
\tablenotetext{d}{We always assumed a power-law and the Galactic absorption. Spectra of
bright sources were analyzed individually, while a photon index of 1.7 was simply
assumed for fainter sources. The Galactic stars would have smaller absorption but
they are not the focus of this work. The four promising candidates for intracluster ULXs
have rest-frame 0.3 - 10 keV luminosities of 0.92 - 6.6 $\times10^{39}$ erg s$^{-1}$,
if in A3627. The other candidates (at least \# 9 and \# 12) have rest-frame 0.3 - 10 keV
luminosities of up to 2.5$\times10^{40}$ erg s$^{-1}$, if in A3627.}
}
\end{center}
\end{table}

\begin{table}
\begin{center}
{\small
\caption{Spectral fits of \gab's regions}
\begin{tabular}{lccc} \hline \hline
Region$^{a}$ & Model$^{b}$ & Parameters$^{c}$ & C-statistic$^{d}$ (d.o.f) \\ \hline

Tail & APEC & $kT$=1.98$^{+0.96}_{-0.56}$, $Z$=0.03$^{+0.43}_{-0.03}$, $L_{\rm 0.5-2 keV}$ (gas)=(3.2$\pm0.5)\times10^{40}$ & 7.6 (10) \\
     & APEC+POW & $kT$=1.08$^{+1.00}_{-0.51}$, $Z$=0$^{+0.23}_{-0}$, $\Gamma$=(1.7) & 6.4 (9) \\
     &          & $L_{\rm 0.5-2 keV}$ (gas)=(2.1$\pm0.5)\times10^{40}$, $L_{\rm 0.3-10 keV}$ (PL)=(3.8$\pm1.0)\times10^{40}$ & \\
     & APEC+POW & $kT$=1.11$^{+1.95}_{-0.27}$, $Z$=(1.0), $\Gamma$=(1.7) & 8.6 (10) \\
     &          & $L_{\rm 0.5-2 keV}$ (gas)=(9.5$\pm3.6)\times10^{39}$, $L_{\rm 0.3-10 keV}$ (PL)=(5.9$\pm1.1)\times10^{40}$ & \\
     & CEMEKL & $\alpha$=1.5$^{+0.7}_{-0.4}$, $kT_{\rm max}$=(6.0), $Z$=(1.0) & 9.2 (11) \\
Nucleus & APEC+PHABS(POW) & $kT$=1.04$^{+0.47}_{-0.26}$, $Z$=0$^{+0.04}_{-0}$, $N_{\rm H}=3.8^{+0.8}_{-0.6}\times10^{23}$, $\Gamma$=(1.7) & 5.1 (11) \\
        & APEC+POW+PHABS(POW) & $kT$=0.36$^{+0.44}_{-0.16}$, $Z$=0.01$^{+3.0}_{-0.01}$, $\Gamma_{1}$=(1.7), $N_{\rm H}=4.8^{+1.1}_{-0.9}\times10^{23}$, $\Gamma_{2}$=(1.7) & 2.3 (10) \\
        &                 & $L_{\rm 0.5-2 keV}$ (gas)=(1.9$\pm0.4)\times10^{40}$, $L_{\rm 0.3-10 keV}$ (nuc.)=(1.7$\pm0.4)\times10^{42}$ & \\

\hline \hline
\end{tabular}
\vspace{-1cm}
\tablenotetext{a}{The nucleus region is defined as a circle with a radius of 5.7 kpc.
The tail region includes all parts of the source except the nuclear region.
Note that the local PSF is not small (see the caption of Figure 13).
We only tried simple models as the statistics are poor.}
\tablenotetext{b}{The Galactic absorption component (1.73$\times10^{21}$ cm$^{-2}$) is included in all cases so any ``PHABS'' component shown is the intrinsic absorption.}
\tablenotetext{c}{The unit of $kT$ is keV and the unit of $Z$ is solar. The unit of luminosity is erg s$^{-1}$ and the unit of absorption is cm$^{-2}$. Parameters in parentheses are fixed.}
\tablenotetext{d}{Check the caption of Table 1 for C-statistic.}
}
\end{center}
\end{table}

\begin{figure}
\vspace{-0.4cm}
\centerline{\includegraphics[height=0.44\linewidth]{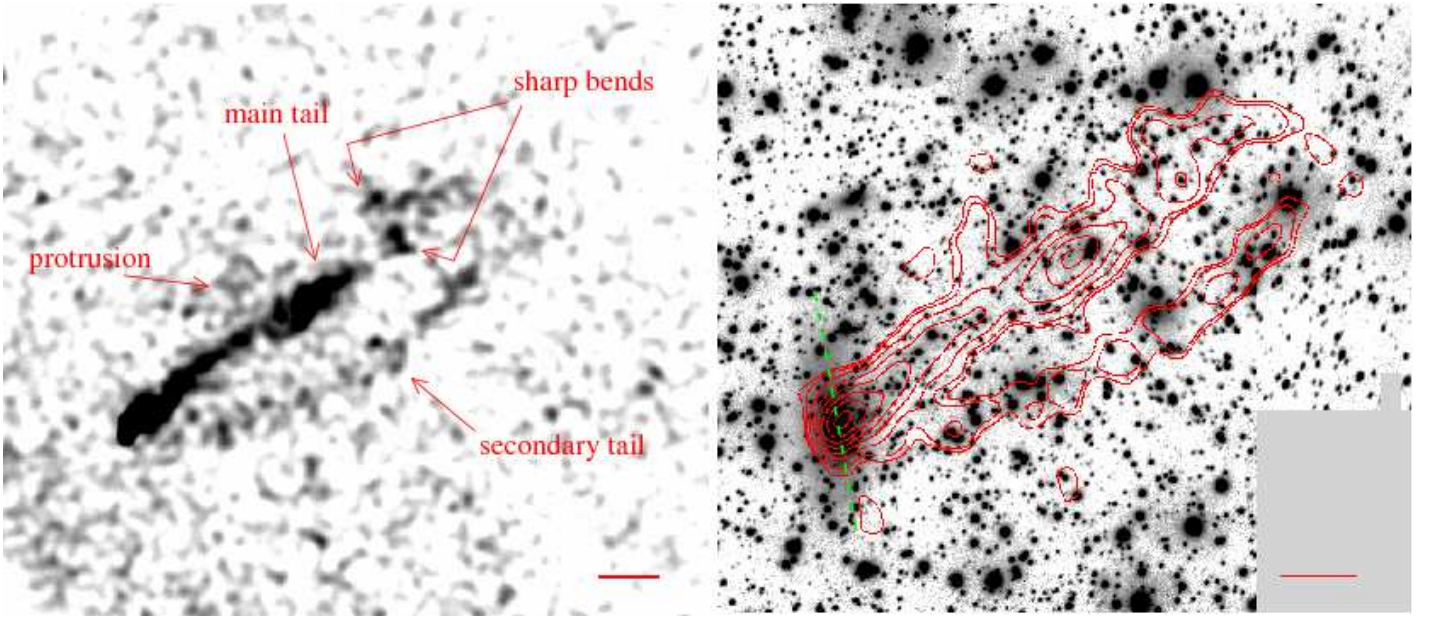}}
  \caption{{\bf Left}: \chandra\ 0.6 - 2.0 keV count image of \ga\ (no exposure
correction) with
main features marked. Point sources were removed and the count image was
smoothed with a 10-pixel (4.92$''$) Gaussian. This image shows the raw data in
a minimally processed way. Besides two significant tails, some sub-structures like
a ``protrusion'' and two sharp bends are clearly visible. The red scale bar
is 10 kpc (or 30.6$''$).
{\bf Right}: the 0.6 - 2.0 keV \chandra\ contours in red superposed
on the SOAR H$\alpha_{\rm on}$ image (H$\alpha$ + continuum). The \chandra\ image
was background subtracted and exposure corrected. Point sources were also
removed. ASMOOTH was used to adaptively smooth the \chandra\ image.
The contours are in square-root spacing and the innermost level is 3 times
the outermost level (note that this image is heavily smoothed so check Figure 3
and 4 for the surface brightness).
The bright part of the H$\alpha$ tail (the first 20 kpc) is also
visible. The green dashed line shows the major axis of the disk plane.
The red scale bar is 10 kpc (or 30.6$''$).
}
\end{figure}

\begin{figure}
\vspace{-0.4cm}
\centerline{\includegraphics[height=0.46\linewidth]{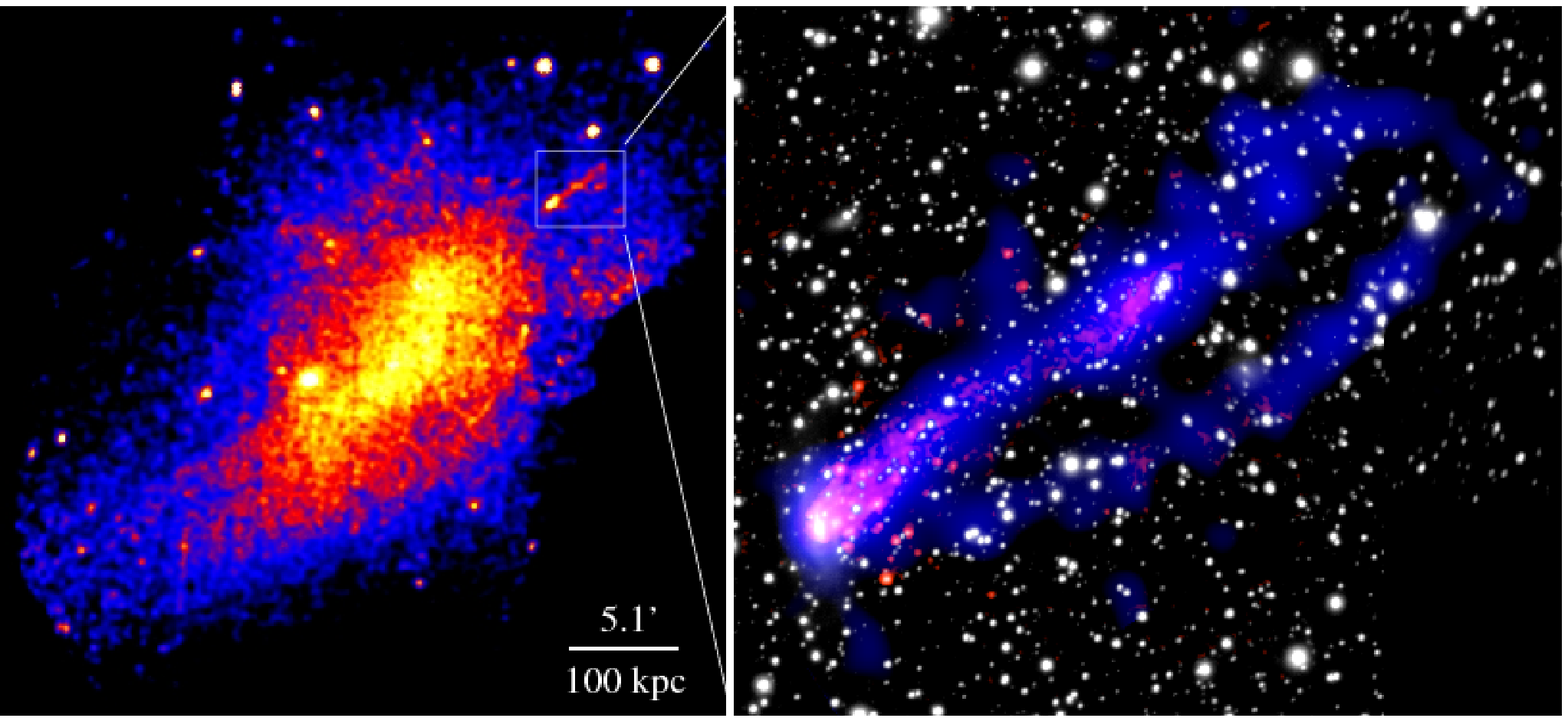}}
  \caption{{\bf Left}: \xmm\ 0.5 - 2 keV mosaic of A3627 from an 18 ks
observation. The main tail of \ga\ is significant in the \xmm\ image.
{\bf Right}: the composite X-ray/optical image of \ga's tail. The \chandra\ 0.6 - 2 keV
image (from the new 140 ks observation) is in blue, while the net H$\alpha$
emission (from the SOAR data, S07) is in red. The white stellar image is
the same as the one shown in Fig. 1. Note that the X-ray image was adaptively
smoothed (also for other X-ray contours shown in this work). The X-ray leading
edge is in fact in the same place as the H$\alpha$ edge (Figure 4).
}
\end{figure}
\clearpage

\begin{figure}
\vspace{-2.5cm}
\centerline{\includegraphics[height=0.8\linewidth]{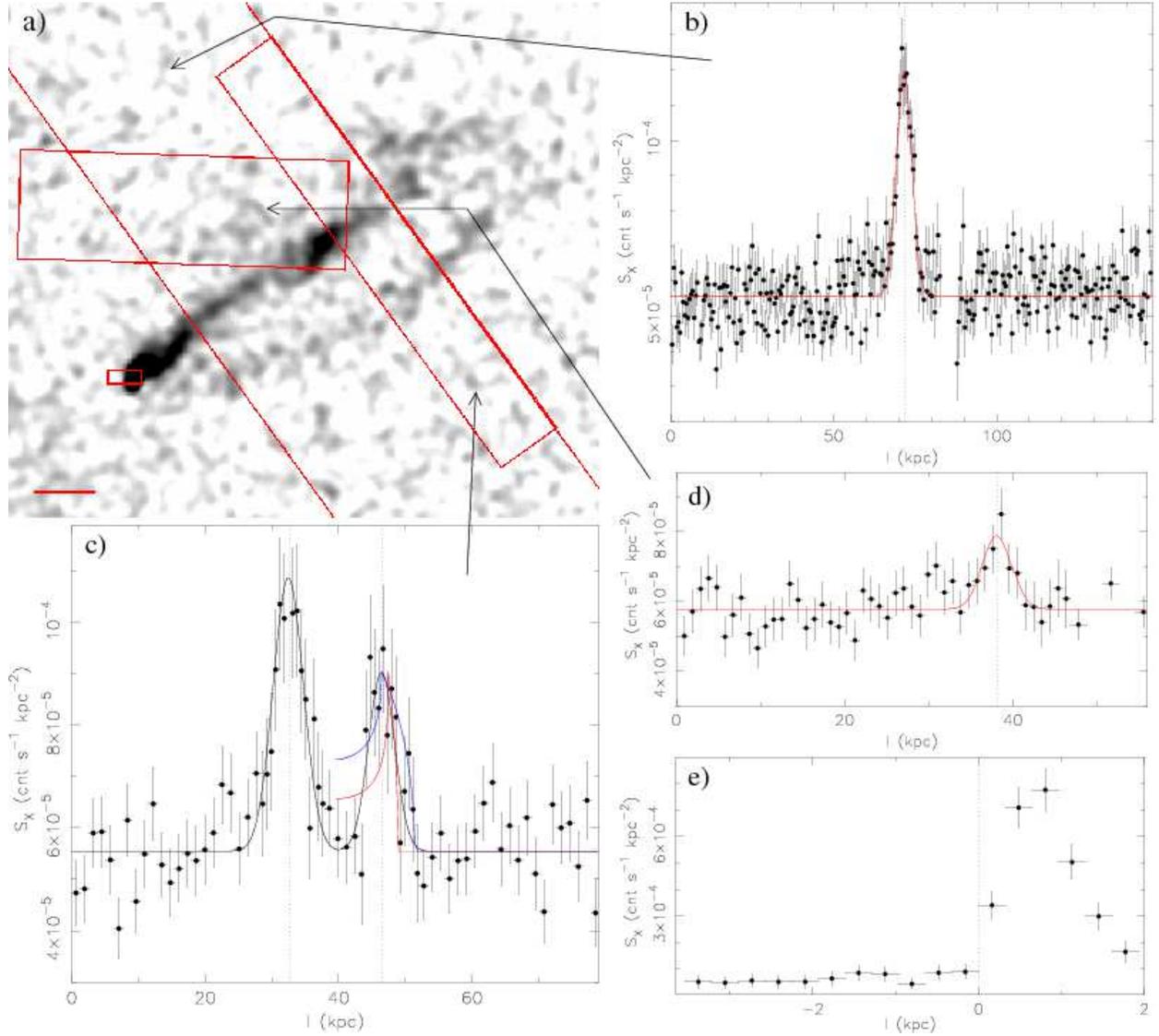}}
  \caption{{\bf a)}: \chandra\ 0.6 - 2.0 keV image (point sources removed,
particle background subtracted and exposure corrected) with four rectangular
regions where the surface brightness profiles are measured (see b - e).
The two unconnected lines are the sides of a rectangle which extends beyond the image.
The red scale bar is 10 kpc (or 30.6$''$).
{\bf b)}: the 0.6 - 2.0 keV surface brightness profile across the straight part
of the main tail.
The width of slices is 115$''$ (or 37.6 kpc). The secondary tail, ``protrusion''
and point sources were masked. A Gaussian fit with a constant background is
also shown. The FWHM of the Gaussian is 5.71$\pm$0.34 kpc.
{\bf c)}: the 0.6 - 2.0 keV surface brightness profile along the direction
perpendicular to the tails at $\sim$ 45 kpc from the galaxy. Point sources are masked.
The width of slices is 35$''$ (or 11.4 kpc). The black solid line is a simple
fit with two Gaussians plus a flat background. The separation between the centers
of the two tails is 14 kpc.
The FWHMs are 5.4$\pm$0.9 kpc and 4.6$\pm$0.9 kpc respectively.
We attempted two fits assuming a cylindrical shell for the whole tail system
(the blue and red solid lines, only for the secondary tail, see Section 3.1).
As can be seen from the red and blue lines, it is impossible to
fit the width of each tail and the faint emission between the tails simultaneously.
{\bf d)}: the 0.6 - 2.0 keV surface brightness profile across the ``protrusion''.
The width of slices is 54.2$''$ (or 17.7 kpc).
The main tail and point sources are masked. A Gaussian fit
with a constant background is also shown. The FWHM of the Gaussian is
4.05$^{+1.68}_{-1.14}$ kpc. The ``protrusion'' is a 4.1 $\sigma$ feature. 
{\bf e)}: the 0.6 - 2.0 keV surface brightness profile across the leading
edge. Upstream of the edge, the surface brightness
is nearly constant at 5.9$\times10^{-5}$ cnt s$^{-1}$ kpc$^{-2}$.
The thickness of the edge is less than 0.2 kpc (or 0.61$''$).
}
\end{figure}
\clearpage

\begin{figure}
\vspace{-0.4cm}
\centerline{\includegraphics[height=0.35\linewidth]{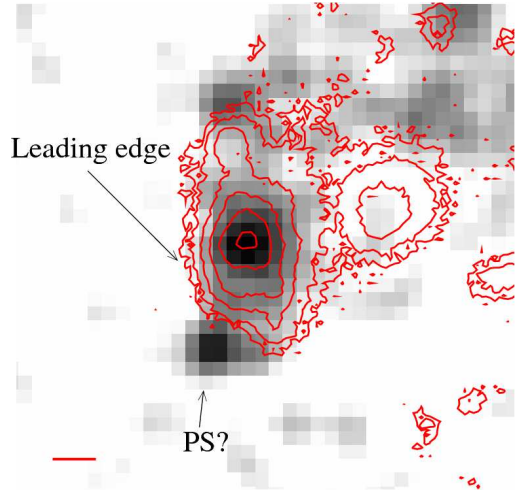}}
  \caption{\chandra\ 0.5 - 7 keV image of \ga's nuclear region superposed on the net H$\alpha$
emission in red contours. The X-ray leading edge is positional coincident with
the H$\alpha$ leading edge. The scale bar is 0.5 kpc (or 1.53$''$).
}
\end{figure}

\begin{figure}
\centerline{\includegraphics[height=0.47\linewidth]{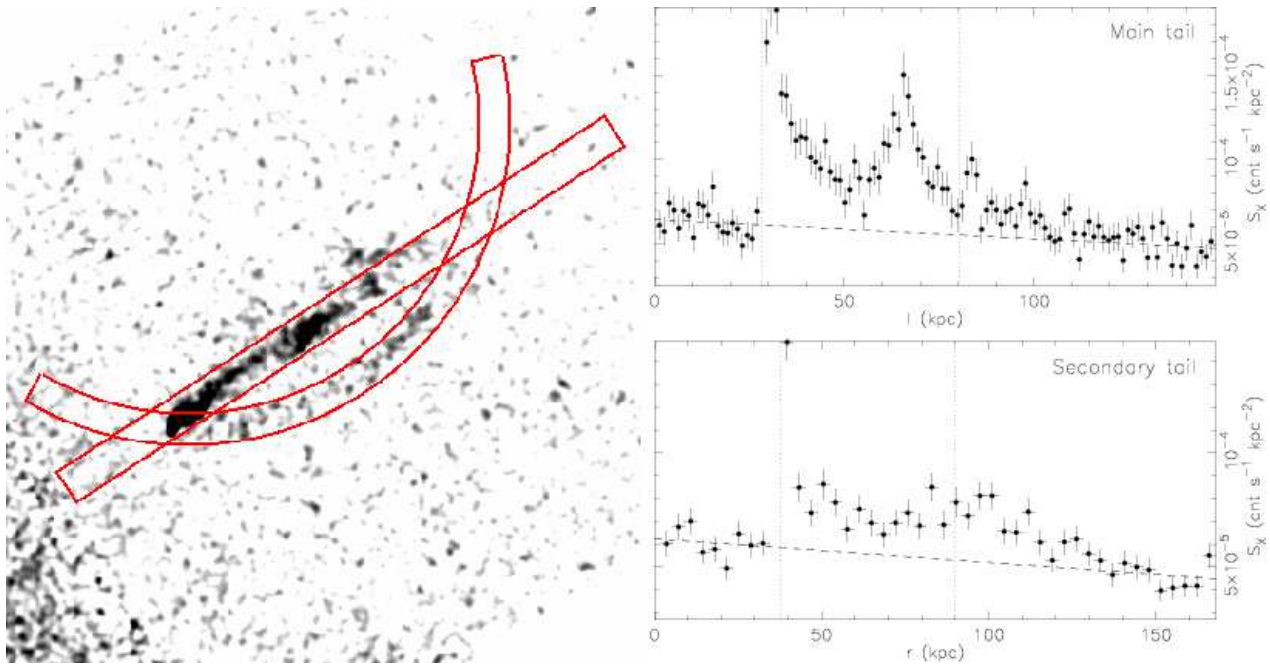}}
  \caption{The 0.6 - 2.0 keV surface brightness profiles along the main
tail and the secondary tail. The regions used to measure the surface brightness
are shown in the left. The width of the regions is 25$''$ for the main tail and
21.4$''$ for the secondary tail. The end of the secondary tail is not curved
as shown (see Figure 1 and 2) but we chose so for simplicity. The dashed lines
represent the local background. The dotted lines in the plot of the main tail
show the positions of the nucleus and the sharp bends. We did not adjust
the width of the region to include the entire bends and ``protrusion''.
Their corresponding positions in the secondary tail are also shown with the
dotted lines. The surface brightness around the nucleus is off scale
(see Figure 3e), as we want to emphasize the faint regions.
}
\end{figure}

\begin{figure}
\vspace{-0.4cm}
\centerline{\includegraphics[height=0.4\linewidth,angle=270]{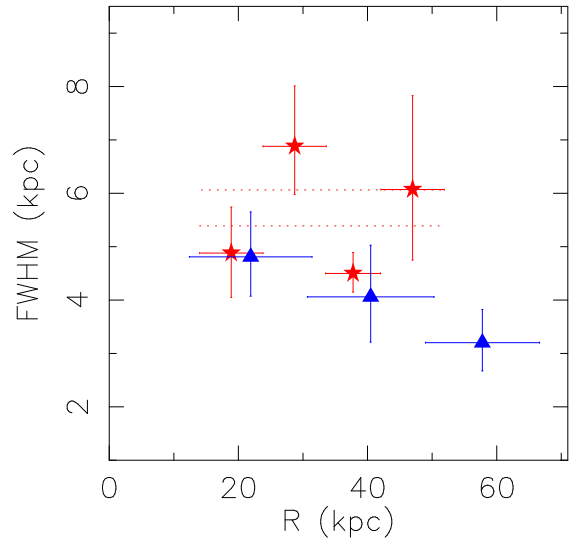}}
  \caption{The FWHM of the main tail (red stars) and the secondary tail (blue
triangles) as a function of the distance from the nucleus. The FWHM values can be taken
as a measure for the width of the tail. For the main tail, we only measured
the width of the straight part, excluding the sharp bends.
Two red dotted lines (5.4 - 6.1 kpc) show the 1 $\sigma$ range of the FWHM
measured for the entire main tail (Figure 3b). As the secondary tail is curved,
the three parts where the width was measured have overlaps. The result shows that
the high density parts of X-ray tails are not expanding.
}
\end{figure}

\begin{figure}
\vspace{-0.4cm}
\centerline{\includegraphics[height=0.51\linewidth]{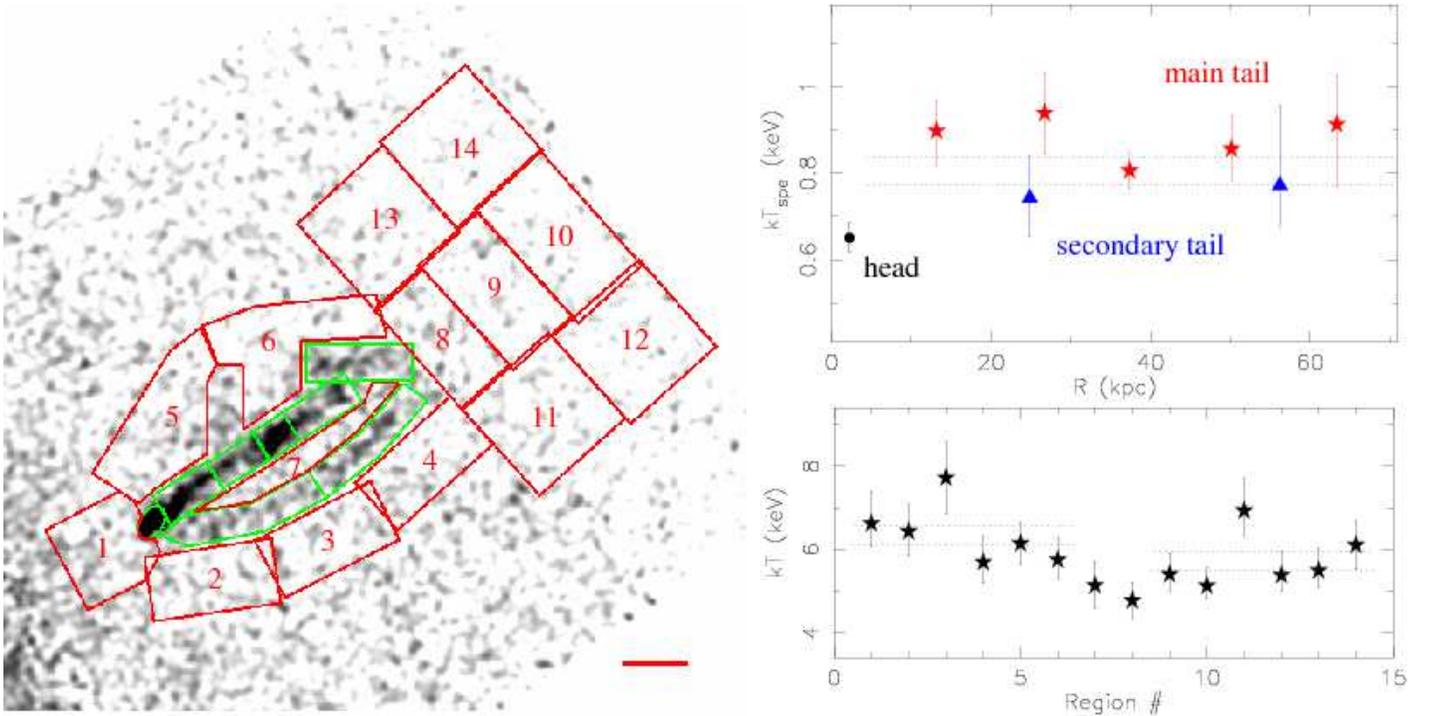}}
  \caption{On the left we show the \chandra\ image with regions where the
ICM or the tail temperatures are measured. The ICM regions are in red, numbered
from 1 to 14 (including a region between two tails, \#7). The galaxy and the
tail regions are in green. We did not measure the temperature of the
``protrusion'' region as it is too faint. The red scale bar is 15 kpc,
which is about the radius of \ga's halo.
On the right we show temperatures of the head, tails and the ICM with
1$\sigma$ errors. We measured temperatures in five regions of the
main tail and two regions of the secondary tail (see the left panel).
These temperatures should be viewed as spectroscopic temperatures for the
multi-phase gas (see Section 3.2 for detail).
The dotted lines show the 1$\sigma$ average temperature range in of the tail
(0.773-0.838 keV), which is consistent with the temperatures of all regions.
Dotted lines in the ICM temperature plot are the 1$\sigma$ average temperature
ranges in the first six regions and the last six regions respectively.
}
\end{figure}

\begin{figure}
\vspace{-0.4cm}
\centerline{\includegraphics[height=0.45\linewidth,angle=270]{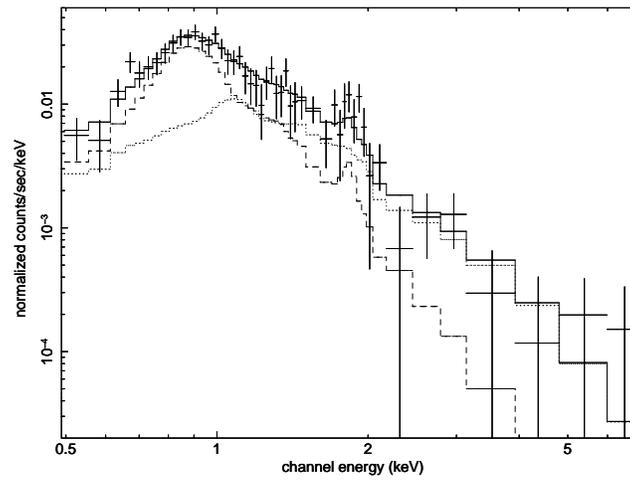}}
  \caption{The \chandra\ spectrum of the whole X-ray tail region of \ga\
(point sources excluded). The best-fit from a 2T model is also shown, with
the dashed and dotted lines representing two components. The iron-L hump and
the Si XIII line are significant. A fit with single thermal component leaves
a significant hard excess.
}
\end{figure}

\begin{figure}
\vspace{-0.4cm}
\centerline{\includegraphics[height=1.0\linewidth]{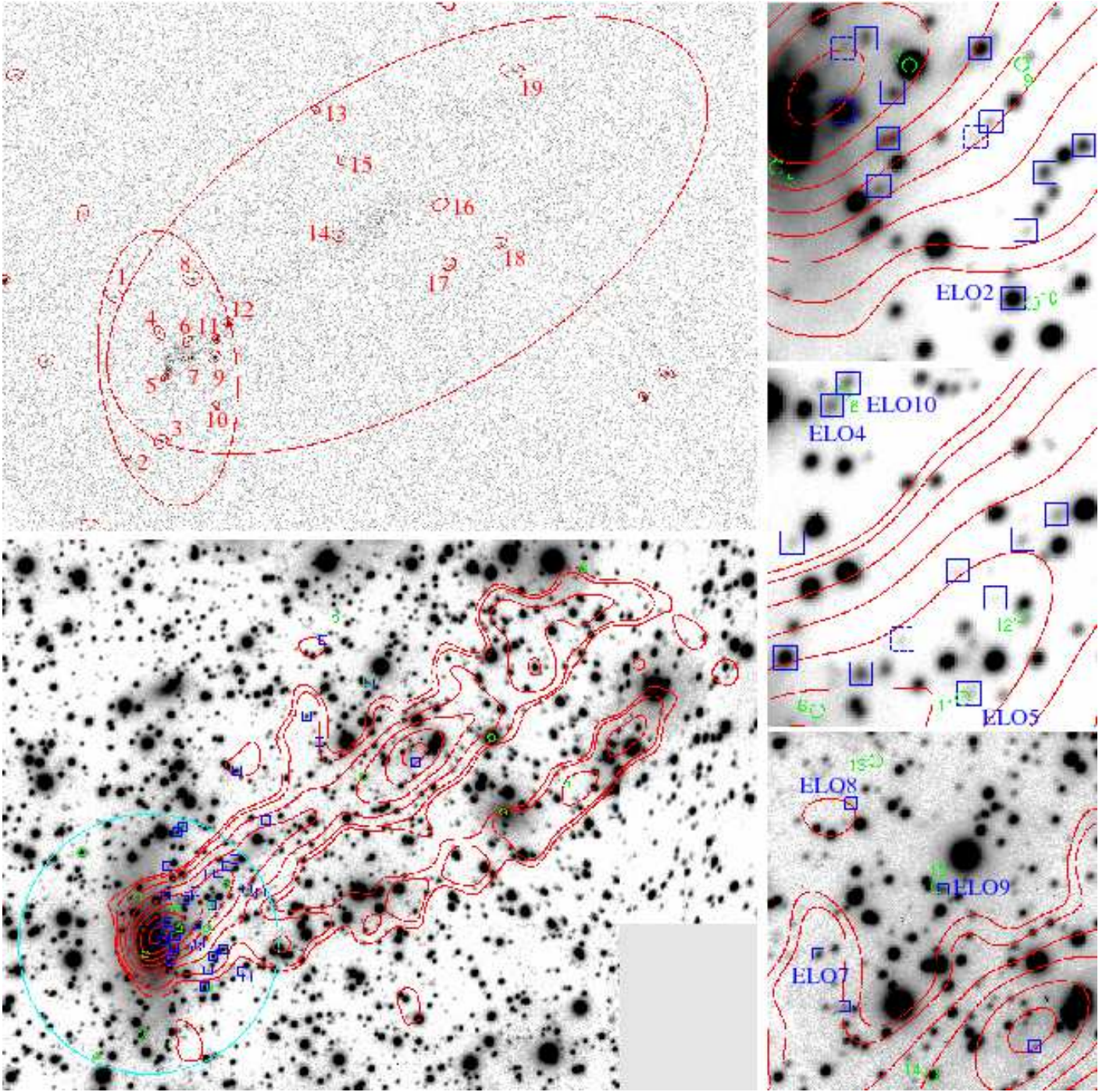}}
  \caption{\chandra\ 0.5 - 7 keV count image is shown in the upper left panel
with point sources marked (Table 3). Two ellipses (10 kpc x 20 kpc and
23.1 kpc x 48.7 kpc respectively, all semi-axes) are the galaxy and tail regions
where we examined the properties of the point sources. The left lower panel shows the
positions of the 19 \chandra\ point sources (green circles) in the two ellipses
and the positions of the 35 HII regions (blue squares). All emission-line objects
identified by S07 with \gem\ spectra (33 in total) are confirmed to be HII regions in A3627.
Two HII region candidates (without \gem\ spectra) are marked with dashed squares.
The cyan circle has a radius of 15 kpc, which is about the size of \ga's
tidally truncated halo.
The right panels show three zoom-in regions. The \chandra\ source number is
also shown. Sources 8, 10, 11 and 15 are promising candidates of intracluster
ULXs, as they are 0.26 - 0.65 kpc from confirmed HII regions and have no
optical counterparts. Sources 9 and 12 are also close to confirmed
HII regions and have no optical counterparts (Table 2). Five \chandra\ sources
have no optical counterparts, including sources 13 and 18 around the tail
(Table 2).
}
\end{figure}

\begin{figure}
\vspace{-0.4cm}
\centerline{\includegraphics[height=0.56\linewidth]{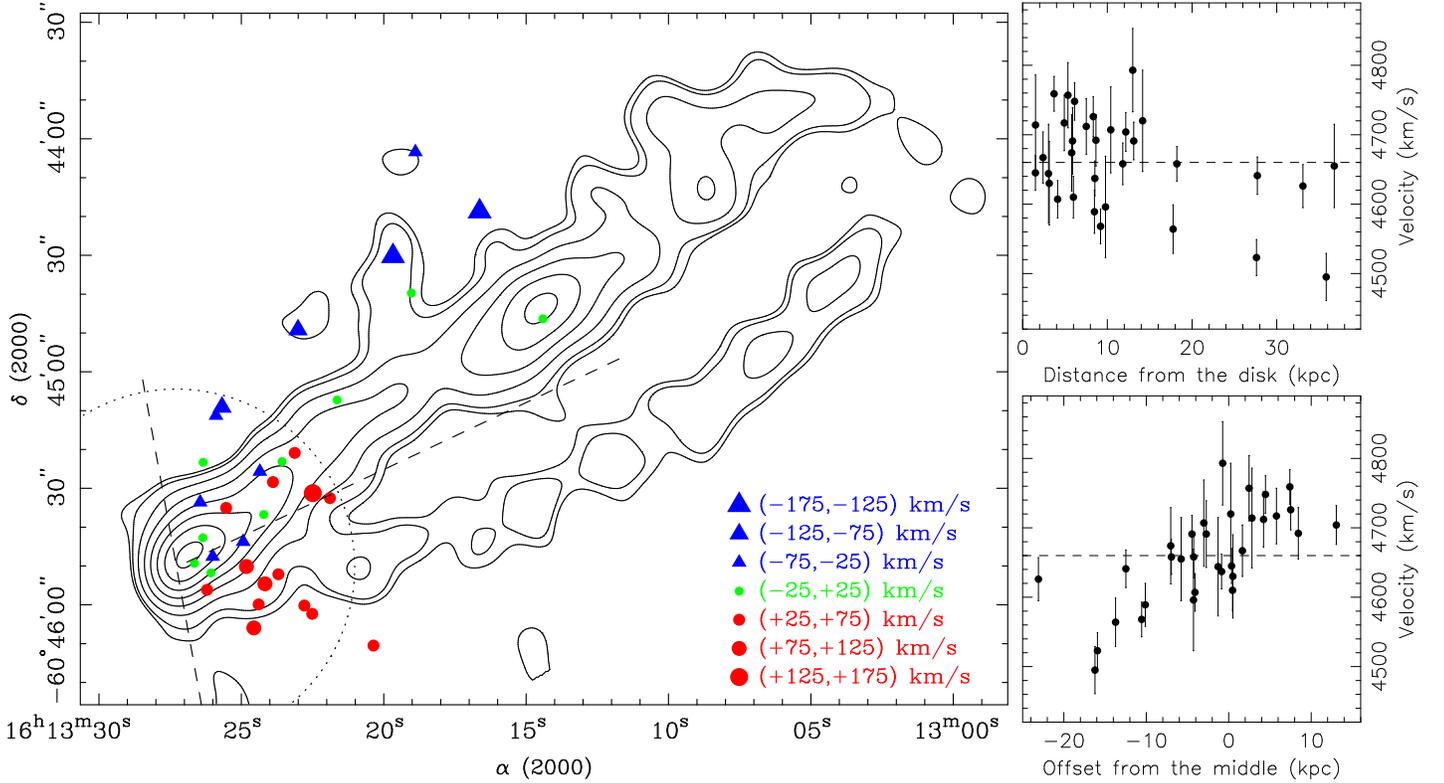}}
  \caption{The heliocentric velocity map of 33 HII regions derived from the
\gem\ data, superposed on the \chandra\ contours, is shown on the left.
Velocities are shown as differences from the system velocity of \ga\
(4660 km/s, see Section 5.2). The imprint of \ga's disk rotation pattern
can be seen in the velocity map (see Section 5.1). The almost vertical dashed
line is along the
major axis of the galaxy, while the other dashed line runs approximately
midway between the two tails, 15 deg from the minor axis of the galaxy.
The dotted circle (with a radius of 15 kpc) is about the size of \ga's tidally
truncated halo. The upper right panel shows the velocities of the HII
regions with the projected distance to the major axis of the galaxy shown in
the left. The dashed line is \ga's system velocity.
The lower right panel shows the velocities with
the projected distance to the middle dashed line between two tails.
Positive offsets are to the south of the dashed line.
For the 26 HII regions within 15 kpc from the projected major axis (likely
still bound), the weighted average velocity is 4671 km/s and the
radial velocity dispersion is 39$^{+13}_{-12}$ km/s. For all 33 HII regions,
the weighted average velocity is 4652 km/s and the
radial velocity dispersion is 54$^{+12}_{-10}$ km/s.
}
\end{figure}

\begin{figure}
\vspace{-0.5cm}
\centerline{\includegraphics[height=1.1\linewidth]{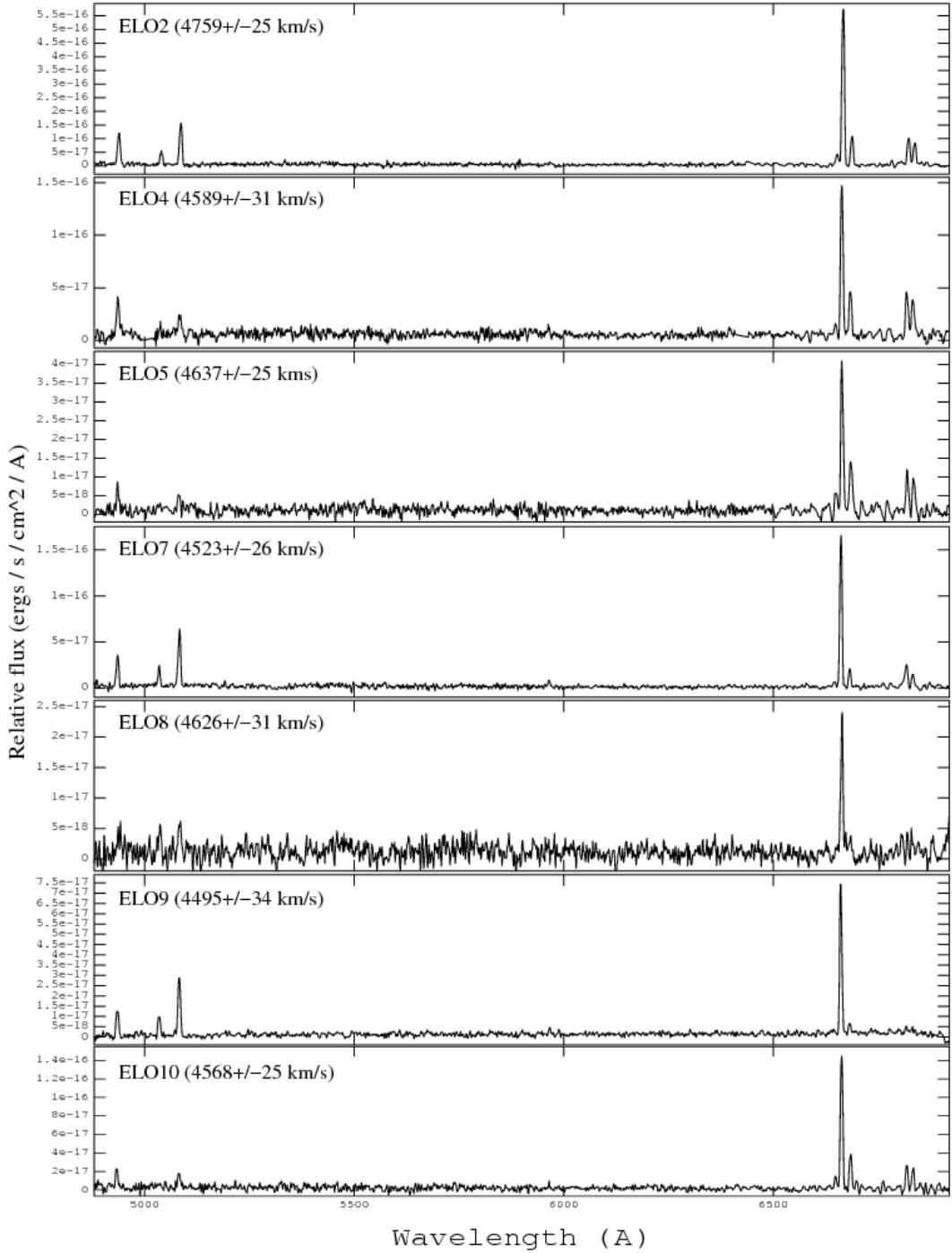}}
  \caption{\gem\ GMOS spectra of seven HII regions, including five sources around
the \chandra\ point sources \#8, \#10, \#11 and \#15. The other two intracluster
HII regions shown (ELO7 and ELO8, see Figure 9), along with ELO9, are among ones
that are farthest from the galaxy (up to 40 kpc). Spectra are calibrated for
the relative flux. Corrections were made for Galactic and atmospheric
extinction. Their spectra are similar to that of the central
emission nebula of \ga\ (Figure 5 of S07, see the labels of lines).
The low [OI] $\lambda$6300 / H$\alpha$ and
[NII] $\lambda$66584 / H$\alpha$ ratios are typical of giant HII regions.
The heliocentric velocities of these HII regions are also shown.
}
\end{figure}

\begin{figure}
\vspace{-0.4cm}
\centerline{\includegraphics[height=0.8\linewidth]{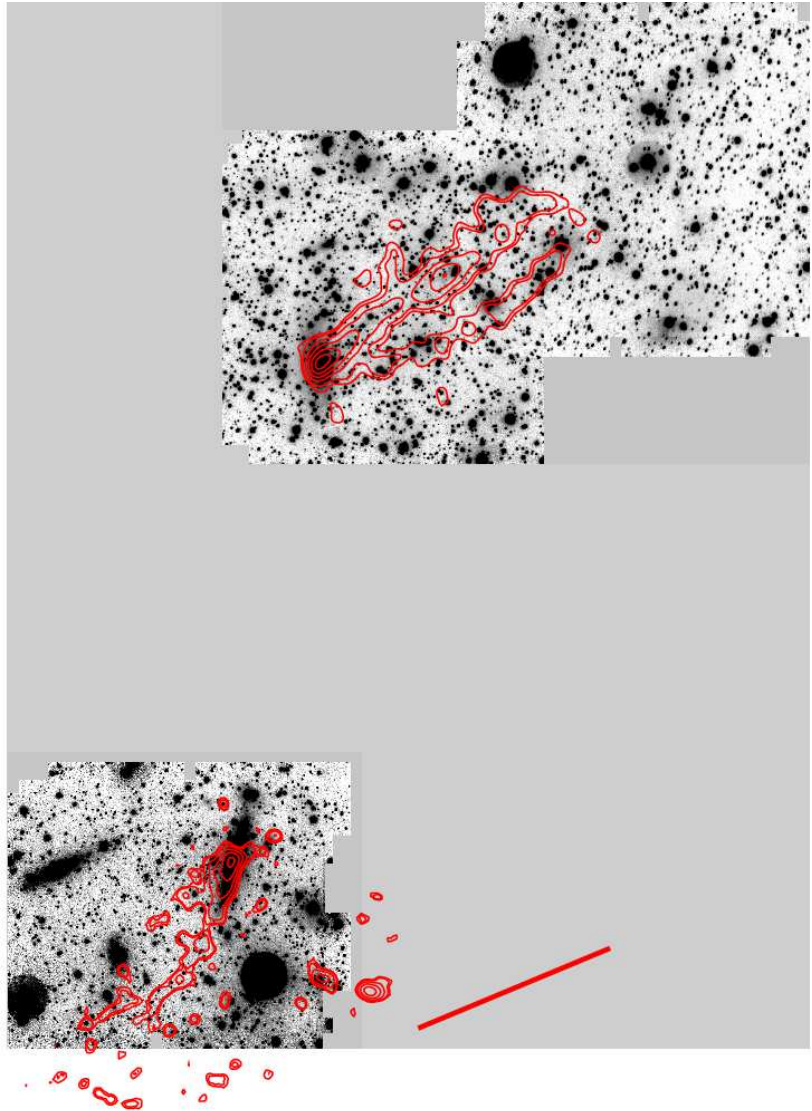}}
  \caption{\chandra\ 0.6 - 2.0 keV contours of \ga's tails and \gab's tail on the SOAR
H$\alpha$+continuum image (in real separation).
The red bar (50 kpc length) shows the position angle of the elongated E/S0 population
within 0.67 Mpc radius (Woudt et al. 2008).
}
\end{figure}
 
\begin{figure} 
\hspace{-0.4cm}
\centerline{\includegraphics[height=0.42\linewidth]{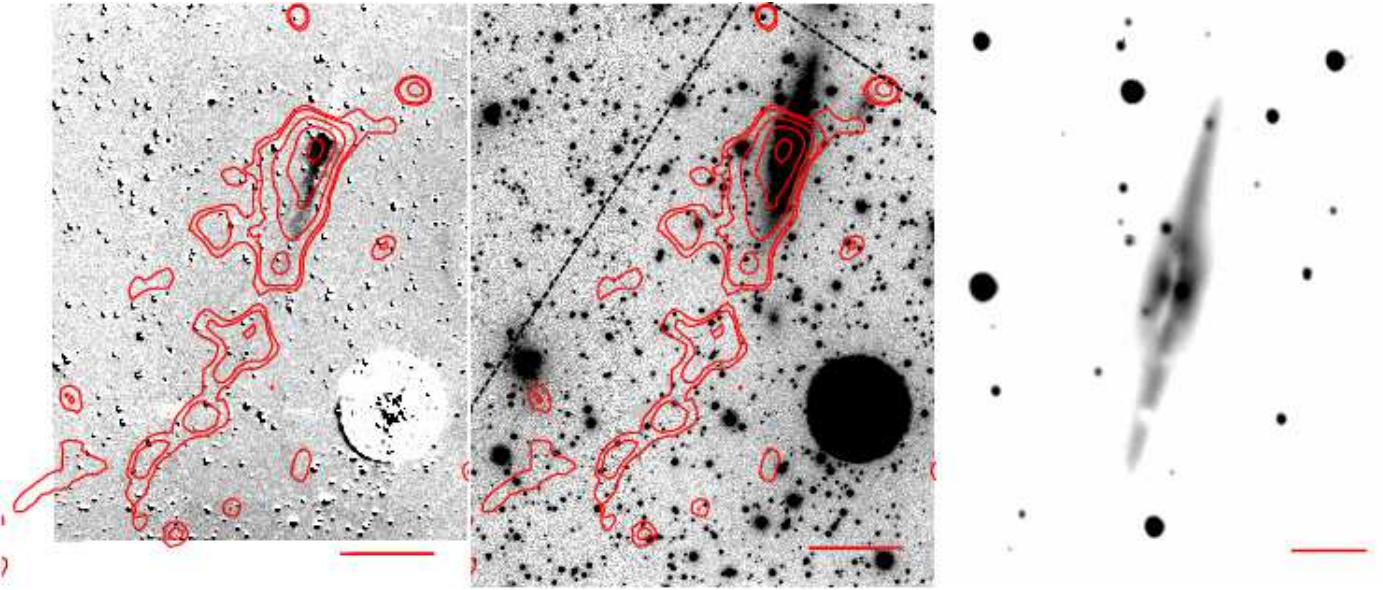}}
  \caption{{\bf Left:} \chandra\ 0.6 - 2.0 keV contours in red superposed on the
net H$\alpha$ emission of \gab. The position of the leading H$\alpha$ edge is 
consistent with that of the X-ray edge. Note that \gab\ is 10.8$'$ off-axis
in this \chandra\ observation so the PSF is significantly blurred (50\% - 90\% encircled
energy radius at 1.49 keV is 7.5$''$ - 14$''$).
H$\alpha$ absorption in the disk plane is visible upstream of the edge.
The H$\alpha$ trail can be traced to at least 20 kpc from the nucleus,
while the X-ray tail can be traced to at least 40 kpc from the nucleus.
Note that the tail region is unfortunately close to a bright star ($V$=10 mag).
The red scale bar is 10 kpc (or 30.6$''$).
{\bf Middle:} \chandra\ 0.6 - 2.0 keV contours in red superposed on the
H$\alpha_{\rm off}$ image to show the galactic emission at low surface
brightness. The black dashed line shows the edge of the I3 chip.
The red scale bar is still 10 kpc (or 30.6$''$).
{\bf Right:} the H$\alpha_{\rm on}$ image of \gab's center shows a clear
dust lane. The red scale bar is 2 kpc (or 6.12$''$).
}
\end{figure}

\begin{figure}
\vspace{-0.4cm}
\centerline{\includegraphics[height=0.42\linewidth]{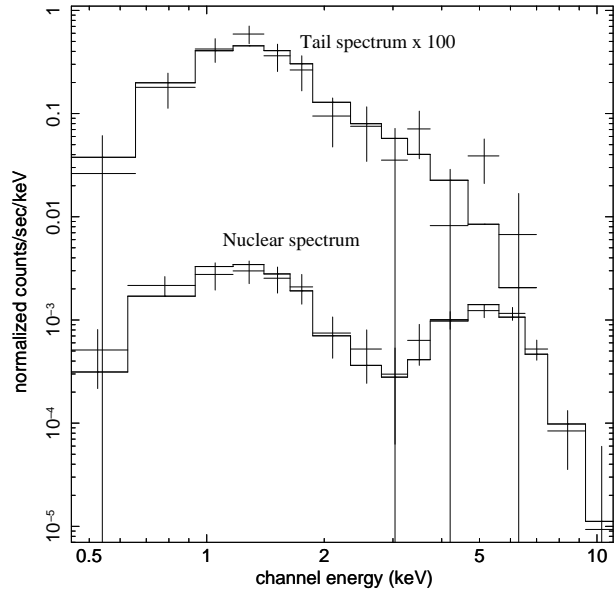}}
  \caption{The \chandra\ spectra of the nuclear and tail regions of \gab.
\gab\ hosts an absorbed AGN. Its X-ray tail appears hotter than \ga's
($\sim$ 2 keV vs. $\sim$ 0.8 keV).
}
\end{figure}

\begin{figure}
\vspace{-0.4cm}
\centerline{\includegraphics[height=0.45\linewidth]{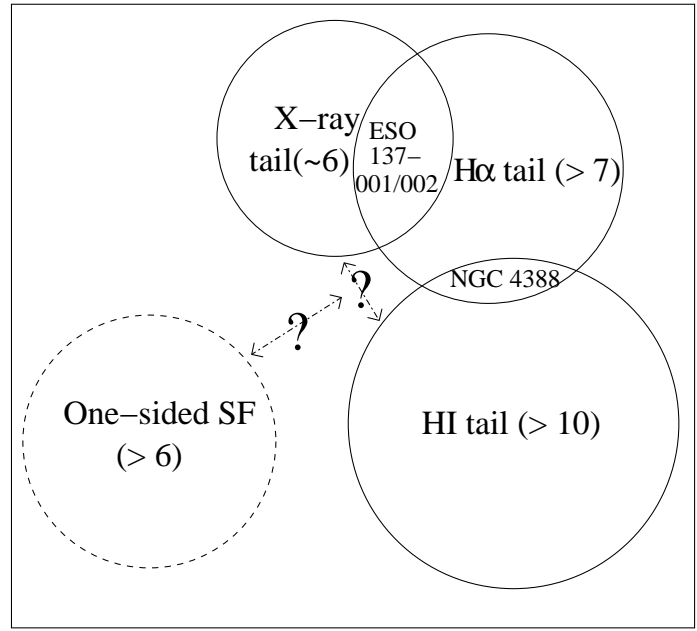}}
  \caption{The Venn diagram of tails of cluster late-type galaxies in HI,
H$\alpha$ and X-rays (see Section 7.1 for detail).
There is little overlap now, mostly because of the lack of data.
Multi-wavelength studies of tails are important to understanding mixing and
other important physical processes. The general connection of one-sided star formation
behind cluster late-type galaxies with ram pressure stripping is also intriguing.
}
\end{figure}

\begin{figure}
\vspace{-1.4cm}
\centerline{\includegraphics[height=0.4\linewidth]{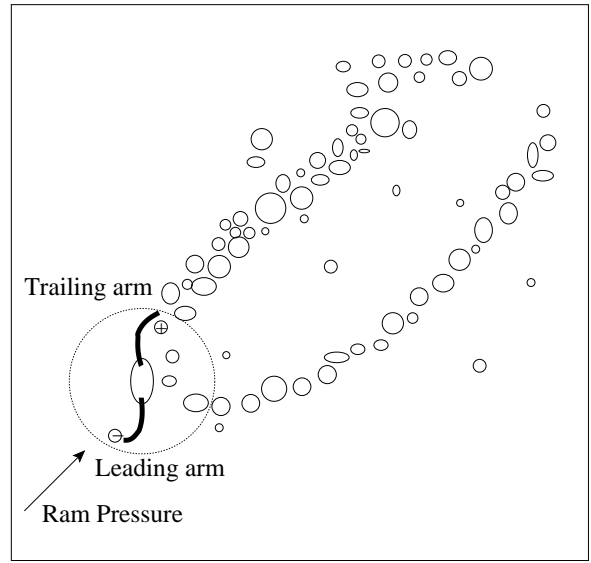}}
  \caption{A cartoon of \ga's stripping event. We assume that the two X-ray
tails come from two major spiral arms, visible in the optical images.
The ISM of the leading arm is more difficult to be stripped, as gas has to
move to deeper potential first. Some of the gas originally around the leading
arm may move backwards to join the tail from the trailing arm. On the contrary,
the ISM around the trailing arm is much easier to be stripped. These factors
explain why the main tail from the trailing arm is brighter.
Both tails can be very clumpy and the data of the cold gas (HI and CO) are
required for a better understanding.
The curvature of the tail can be explained by the galactic rotation.
HII regions are distributed in a larger opening angle (Fig. 10) as they are decoupled from
the ram pressure.
}
\end{figure}

\end{document}